\documentclass[10pt,aps,pra,twocolumn,floatfix]{revtex4-2}
\usepackage{amsthm}
\usepackage{graphicx}
\usepackage{dcolumn}
\usepackage{bm}
 \usepackage{amsmath}
\usepackage{physics}
\usepackage{xcolor}
\newcommand{\dif}{\mathop{}\!\mathrm{d}}

\newtheorem{definition}{Definition}
\newtheorem{remark}{Remark}

\usepackage[caption=false]{subfig}
\usepackage{ragged2e}

\newtheorem{lemma}{Lemma}
\newtheorem{theorem}{Theorem}
\usepackage{indentfirst}
\usepackage{titlesec}
\usepackage{amssymb}
\titlespacing*{\section}{0pt}{*1.2}{*0.8}       
\titlespacing*{\subsection}{0pt}{*1.0}{*0.6}

\begin{document}

\preprint{APS/123-QED}

\title{Two-stage Quantum Estimation and Asymptotics of Quantum-enhanced Transmittance Sensing}

\author{Zihao Gong}
\altaffiliation{Present address: Department of Electrical and Computer Engineering, University of Maryland, College Park, MD 20742, USA}
\email{zgong12@umd.edu}
\author{Boulat A. Bash}
\email{boulat@arizona.edu}
\affiliation{Electrical and Computer Engineering Department, University of Arizona, Tucson, AZ 85721, USA}

\begin{abstract}
We consider the estimation of a single unknown parameter embedded in a quantum state.
Quantum Cram\'{e}r-Rao bound (QCRB) is the ultimate limit of the mean squared error for any unbiased estimator. 
While it can be achieved asymptotically for a large number of quantum state copies, the measurement required often depends on the true value of the parameter of interest. 
Prior work addresses this paradox using a two-stage approach: in the first stage, a preliminary estimate is obtained by applying, on a vanishing fraction of quantum state copies, a sub-optimal measurement that does not depend on the parameter of interest.
In the second stage, the preliminary estimate is used to construct the QCRB-achieving measurement, which is then applied to the remaining quantum state copies.
This is akin to two-step estimators for classical problems with nuisance parameters.
Unfortunately, the original analysis imposes conditions that severely restrict the class of classical estimators applicable to quantum measurement outcomes, hindering the application of this method. 
We relax these conditions to substantially broaden the class of usable estimators for single-parameter problems at the cost of slightly weakening the asymptotic properties of the two-stage method. 
We also account for nuisance parameters.
We apply our results to obtain the asymptotics of quantum-enhanced transmittance sensing.

\end{abstract}

\maketitle

\section{Introduction} \label{sec:introduction}
Consider the estimation of a scalar parameter $\theta$ embedded in a quantum state $\hat{\sigma}(\theta)$ of a physical system. 
Quantum mechanics governs precision of this process \cite{helstrom76quantumdetect}:
the quantum Cram\'{e}r-Rao bound (QCRB) is the ultimate limit on the minimum mean squared error (MSE) of an unbiased estimator $\check{\theta}$ applied to the outcomes of an arbitrary quantum measurement of $\hat{\sigma}(\theta)$.
The search for physical measurement devices that achieve QCRBs in various domains is a central problem in quantum metrology and sensing \cite{giovannetti11metrology, degen17quantumsensing}.
Success in this will not only advance fundamental physics but also impact our everyday lives \cite{ye24quantumsensing}.
Indeed, even classical devices can be substantially improved by applying quantum-inspired analysis (e.g., optical sub-diffraction-limit resolution using spatial mode sorting \cite{tsang16superresolution, tsang19superresolutionsurvey}). 

Fortunately, for scalar estimation problems, quantum estimation theory provides a ``turn-the-crank'' method to derive a mathematical expression for the QCRB-achieving measurement (see \cite[Ch.~VIII.4(b)]{helstrom76quantumdetect} and the summary in Section \ref{sec:quantum estimation prerequisites}).
Unfortunately, this optimal measurement structure may 1) lack a physical realization; and 2) depend on the true value of the parameter to be estimated.
The first issue is specific to a particular sensing problem.
Here, we focus on the paradox raised by the second issue.
Given $n\to\infty$ copies of $\hat{\sigma}(\theta)$, it can be resolved by a sequential method that first randomly guesses $\theta$ to construct the measurement for the first copy of $\hat{\sigma}(\theta)$.
A measurement for each subsequent copy of $\hat{\sigma}(\theta)$ is built from the previous estimate of $\theta$, refining the estimator by evolving it towards the optimal  \cite{nagaoka1988asymptotically, fujiwara2006strong}. 
Under certain regularity conditions, this technique yields strongly consistent and asymptotically normal  estimators of $\theta$ \cite{fujiwara2006strong}. 
However, repeated adjustments to the quantum-measurement device required by sequential estimation are often impractical. 
This motivates the two-stage method \cite{gill00twostagemeasurement, hayashi2005statistical}, \cite[Ch.~6.4]{hayashi17qit}: in the preliminary stage, a suboptimal measurement independent of $\theta$ is applied to a fraction of states that diminishes with $n$. This estimate is used to construct the optimal measurement and refine the estimate in the second, refinement stage. 
The two-stage approach is similar to the two-step method used in classical statistics, which estimates nuisance parameters in the first step, and uses the result to estimate the parameter of interest in the second step \cite[Sec.~6]{newey94largesampleestimation}. 
However, its analysis typically focuses on the impact of errors in the first step on the consistency and efficiency of the estimator in the second step.
Unfortunately, this does not apply directly to two-stage quantum estimation.

The authors of \cite{hayashi2005statistical}, \cite[Ch.~6.4]{hayashi17qit} were the first to present a comprehensive asymptotic treatment of the two-stage method in the context of quantum sensing.
They show that, under certain regularity conditions, the normalized MSE of the two-stage estimator approaches the QCRB as $n\to\infty$ (we re-state this formally as Lemma \ref{lemma:HM} in Section \ref{sec:prior work}).
Arguably, this is the strongest result one can expect for any estimator. 
Unfortunately, its applicability is limited by the stringency of the regularity conditions imposed on the classical estimators that process the outcomes of quantum measurements.
Thus, the first of our two contributions is the relaxation of the regularity conditions to allow asymptotic analysis of a substantially larger class of estimators, including many maximum likelihood estimators (MLEs).
The cost is a slight weakening of the asymptotic properties: we show that, like MLE \cite{kay93statSP1, vantrees01part1, poor94intro-det-est}, the two-stage estimator under our conditions is consistent and asymptotically normal, with the variance of the limiting Gaussian matching QCRB.
However, these weakened properties suffice for practical tasks such as estimating the confidence intervals \cite[Sec.~1]{newey94largesampleestimation}.
Our Theorem \ref{theorem: our} in Section \ref{sec:main} states the conditions for both weak and strong consistency. While we presented these in the preliminary conference version of this work \cite{gong24twostage-isit}, we add a nuisance-parameter analysis here.

Our results enable asymptotic analysis of quantum estimators for many operationally important problems.
For example, they immediately strengthen the existing work on two-stage quantum-inspired sub-diffraction-limit imaging \cite{grace20adaptivesuperresolution, Sajjad21, Sajjad24, ozer2024adaptivesuperresolutionimagingprior, deshler2025quantumlimitedspatialresolution}\footnote{We defer the  detailed asymptotic analysis of two-stage quantum-inspired sub-diffraction-limit imaging to future work. However, we note that the observations in \cite{grace20adaptivesuperresolution, Sajjad21, Sajjad24, ozer2024adaptivesuperresolutionimagingprior, deshler2025quantumlimitedspatialresolution} lie in various compact sets, which significantly simplifies the analysis. On the contrary, our analysis in Section \ref{sec:transmittance sensing} is complicated by the observations of photon counts being natural numbers that do not lie in a compact set.}
Furthermore, we can employ our relaxed regularity conditions to obtain the asymptotic behavior of quantum-enhanced power-transmittance sensing in the bosonic channel.
This translates to estimating power loss in many practical channels, including optical, radio-frequency, and microwave.
Previously, we derived the QCRB in \cite{gong21losssensing} and the optimal quantum measurement structure that achieves it in \cite{gong22losssensing}.
The optimal measurement depends on the true transmittance, necessitating a two-stage approach that we numerically analyzed in \cite{gong22losssensing}.
Here, we add an unknown phase offset to the bosonic channel model, which is a nuisance parameter that must also be estimated in the first stage. 
Our Theorem \ref{theorem: our} allows us to show strong consistency and asymptotic normality of estimating power transmittance using the optimal quantum measurement even with the additional unknown phase offset.
This is our second contribution.
Conference version \cite{gong24twostage-isit} of this paper includes preliminary analysis that neither addresses the phase offset nor includes detailed proofs.

This paper is organized as follows: after a brief review of quantum estimation theory in Section \ref{sec:quantum estimation prerequisites}, we formally introduce the two-stage method in Section \ref{sec:twostage}.
We then summarize the results on its asymptotics from \cite{hayashi2005statistical}, \cite[Ch.~6.4]{hayashi17qit}, adapting them to single-parameter quantum estimation of interest to us in Section \ref{sec:prior work}.
Our main results are in Sections \ref{sec:main} and \ref{sec:transmittance sensing}.
We state our relaxed regularity conditions for consistency and asymptotic normality of two-stage estimation as Theorem \ref{theorem: our} in Section \ref{sec:main}.
We then show in Section \ref{sec:transmittance sensing} that the strong form of the conditions for Theorem \ref{theorem: our} holds for the quantum-enhanced transmittance sensor derived in \cite{gong22losssensing}.
We conclude with a discussion of future work in Section \ref{sec: conclusion}.
Appendix contains proofs and derivations.

\section{Two-stage Quantum Estimation}

\subsection{Quantum Estimation Prerequisites} \label{sec:quantum estimation prerequisites}
Let's review the principles and fundamental limits of quantum estimation.
We encourage the reader to consult \cite{helstrom76quantumdetect} for details and proofs.
Denoting by $\Gamma$ the parameter space, we are interested in estimating an unknown parameter $\theta\in\Gamma$ that is physically encoded in a quantum state $\hat{\sigma}(\theta)$.
A positive operator-valued measure (POVM) $\{\hat{A}_x\}$ describes a physical device that extracts information about $\theta$ from $\hat{\sigma}(\theta)$.
POVM is non-negative and complete: $\forall x: \hat{A}_x\succeq0$ and $\sum_x \hat{A}_x = \hat{I}$, where $\hat{I}$ is the identity.
A random variable $X(\theta)$ with probability mass function (p.m.f.) $p_{X(\theta)}(x;\theta) = \trace\{\hat{A}_x\hat{\sigma}(\theta)\}$ describes the classical statistics of the output when a device characterized by POVM $\{\hat{A}_x\}$ is used to measure $\hat{\sigma}(\theta)$ \cite[Ch.~III]{helstrom76quantumdetect}.
 
Given an observed output $x$ from POVM, we desire an unbiased estimator $\check{\theta}(x)$, i.e., $E_{X(\theta_{\mathrm{t}})}\left[\check{\theta}\left(X(\theta_{\mathrm{t}})\right)\right]=\theta_{\mathrm{t}}$, that minimizes the mean square error (MSE)  $V_{\theta_{\mathrm{t}}}\left(\check{\theta}\right)=E_{X(\theta_{\mathrm{t}})}\left[\left(\check{\theta}\left(X(\theta_{\mathrm{t}})\right)-\theta_{\mathrm{t}}\right)^2\right]$, where $\theta_{\mathrm{t}}$ is the true value of $\theta$ and $E_{X(\theta_{\mathrm{t}})}[f\left(X(\theta_{\mathrm{t}})\right)]$ is the expected value of $f\left(X(\theta_{\mathrm{t}})\right)$.  The lower bound on the MSE is the classical Cram\'{e}r-Rao bound (CCRB) \cite{kay93statSP1, vantrees01part1}:
\begin{align}\label{eq:cCRB}V_{\theta_{\mathrm{t}}}\left(\check{\theta}\right)&\ge\mathcal{I}_{\theta}\left(X(\theta_{\mathrm{t}})\right)^{-1},
 \end{align}
where the classical Fisher information (FI) associated with $\theta$ for random variable $X(\theta_{\mathrm{t}})$ is
\begin{align}\mathcal{I}_{\theta}\left(X(\theta_{\mathrm{t}})\right)&=E_{X(\theta_{\mathrm{t}})}\left[\left.\left(\partial_{\theta} \log p_{X(\theta)}(X(\theta_{\mathrm{t}});\theta)\right)^2\right|_{\theta=\theta_{\mathrm{t}}} \right].
\end{align}
and $\partial_{x}f(x,y)=\frac{\partial f(x,y)}{\partial x}$ denotes a partial derivative. Classical FI is additive: for a sequence of $n$ independent and identically distributed (i.i.d.) random variables $\left\{X_k(\theta_{\mathrm{t}})\right\}_{k=1}^n$, $\mathcal{I}_{\theta}\left(\left\{X_k(\theta_{\mathrm{t}})\right\}_{k=1}^n\right)=n\mathcal{I}_{\theta}\left(X_1(\theta_{\mathrm{t}})\right)$. 

Quantum estimation theory allows optimization of POVM $\{\hat{A}_x\}$ that is implicitly fixed in the classical analysis \cite[Ch.~VIII]{helstrom76quantumdetect}, yielding the quantum Cram\'{e}r-Rao bound (QCRB): 
\begin{align}\label{eq:qCRB}V_{\theta_{\mathrm{t}}}\left(\check{\theta}\right)&\ge\mathcal{I}_{\theta}\left(X(\theta_{\mathrm{t}})\right)^{-1}\ge\mathcal{J}_{\theta}\left(\hat{\sigma}(\theta_{\mathrm{t}})\right)^{-1},
\end{align}
where $\mathcal{J}_{\theta}\left(\hat{\sigma}(\theta_{\mathrm{t}})\right) = \trace\left\{\left(\hat{\Lambda}(\theta_{\mathrm{t}})\right)^2\hat{\sigma}(\theta_{\mathrm{t}})\right\}$ is the quantum FI associated with $\theta$ for state $\hat{\sigma}(\theta_{\mathrm{t}})$ and  $\hat{\Lambda}(\theta)$ is the symmetric logarithm derivative (SLD) operator. SLD is Hermitian but not necessarily positive and is defined implicitly by \cite[Ch.~VIII.4(b)]{helstrom76quantumdetect}: 
\begin{align}
\label{eq:sld_def}  \partial_{\theta}\hat{\sigma}(\theta) = \left(\hat{\Lambda}(\theta)\hat{\sigma}(\theta)+\hat{\sigma}(\theta)\hat{\Lambda}(\theta)\right)/2.
\end{align}
Analogous to classical FI, quantum FI is additive: for a tensor product of $n$ states $\hat{\sigma}^{\otimes n}(\theta_{\mathrm{t}})$, $\mathcal{J}_{\theta}\left(\hat{\sigma}^{\otimes n}(\theta_{\mathrm{t}})\right)=n\mathcal{J}_{\theta}\left(\hat{\sigma}(\theta_{\mathrm{t}})\right)$.

Consider a POVM $\mathcal{M}(\theta)=\left\{\ket{\lambda_x(\theta)}\bra{\lambda_x(\theta)}\right\}$ that is constructed from an eigendecomposition of SLD $\hat{\Lambda}_{\theta}=\sum_x\lambda_x(\theta)\ket{\lambda_x(\theta)}\bra{\lambda_x(\theta)}$, where $\left\{\ket{\lambda_x(\theta)}\right\}$ is a set of orthonormal pure eigen-states of $\hat{\Lambda}_{\theta}$ and $\left\{\lambda_x(\theta)\right\}$ are the corresponding eigenvalues.
Note that $\mathcal{M}(\theta)$ depends structurally on the parameter of interest $\theta$; however, it is distinct from the quantum state $\hat{\sigma}(\theta)$ that carries information about $\theta$.
Since $\theta$ in $\mathcal{M}(\theta)$ can be set differently than $\theta$ in $\hat{\sigma}(\theta)$, we describe the outcome statistics of measuring $\hat{\sigma}(\theta)$ using $\mathcal{M}(\theta^\prime)$ by a random variable $X(\theta,\theta^\prime)$ with probability mass function (p.m.f.) $p_{X(\theta,\theta^\prime)}(x;\theta,\theta^\prime) = \trace\left\{\ket{\lambda_x(\theta^\prime)}\bra{\lambda_x(\theta^\prime)}\hat{\sigma}(\theta)\right\}$.
In the following, we denote $\mathcal{J}_{\theta_{\mathrm{t}}}\equiv \mathcal{J}_{\theta}(\hat{\sigma}(\theta_{\mathrm{t}}))$ and $\mathcal{I}_{\theta_{\mathrm{t}},\theta^\prime}\equiv\mathcal{I}_{\theta}\left(X(\theta_{\mathrm{t}},\theta^\prime)\right)$ for brevity.
Measurement $\mathcal{M}(\theta)$ is optimal in the sense that the classical FI in its outcomes equals the quantum FI when it is parameterized by the true value $\theta_{\mathrm{t}}$ of the parameter $\theta$: $\mathcal{I}_{\theta_{\mathrm{t}},\theta_{\mathrm{t}}}=\mathcal{J}_{\theta_{\mathrm{t}}}$.
An efficient estimator extracts the value of $\theta$ from the classical outcomes of this measurement with minimal MSE.
However, knowledge of the true value $\theta_{\mathrm{t}}$ of the parameter $\theta$ is needed to construct the optimal measurement $\mathcal{M}(\theta_{\mathrm{t}})$.

Finally, observed state may depend on $u$ unknown nuisance parameters $\vec{\vartheta}=\left\{\vartheta_i\right\}_{i=1}^u$, where $\vartheta_i\in\Gamma_i$ and $u<\infty$ that does not increase with the number of observations $n$. SLD-eigendecomposition POVM $\mathcal{M}\left(\theta^\prime; \vec{\vartheta}\right)$ and the classical FI in its outcomes $\mathcal{I}_{\theta_{\mathrm{t}},\theta^\prime,\vec{\vartheta}}$ then depend on $\vec{\vartheta}$.  Thus, optimal measurement $\mathcal{M}\left(\theta_{\mathrm{t}}; \vec{\vartheta}_{\mathrm{t}}\right)$ is also parameterized by the true values $\vec{\vartheta}_{\mathrm{t}}$ of $\vec{\vartheta}$, and $\mathcal{I}_{\theta_{\mathrm{t}},\theta_{\mathrm{t}},\vec{\vartheta}_{\mathrm{t}}}=\mathcal{J}_{\theta_{\mathrm{t}}}$. However, we emphasize that we are interested in accurately estimating only the parameter of interest $\theta$ and not the nuisance parameters (otherwise a multi-parameter estimation problem emerges). While we address nuisance parameters at various points in this paper, we omit them from our expressions when appropriate for brevity.

\subsection{Two-stage Quantum Estimator}
\label{sec:twostage}

Adaptive approaches \cite{nagaoka1988asymptotically, fujiwara2006strong, gill00twostagemeasurement, hayashi2005statistical}, \cite[Ch.~6.4]{hayashi17qit} 
resolve the paradox outlined above.
Methods that update the measurement after measuring each state are analyzed in \cite{nagaoka1988asymptotically, fujiwara2006strong}.
Here we focus on the asymptotics of the simpler two-stage approach \cite{gill00twostagemeasurement, hayashi2005statistical}, \cite[Ch.~6.4]{hayashi17qit}.
First, we pre-estimate $\check{\theta}_{\rm p}$ from the first $f(n)\in \omega(1)\cap o(n)$ available states using a sub-optimal measurement that does not depend on $\theta$, where $\omega(1)$ and $o(n)$ denote the respective sets of functions that are asymptotically larger than a constant and smaller than $n$.  That is, $\lim_{n\to\infty}f(n)=\infty$ and $\lim_{n\to\infty}\frac{f(n)}{n}=0$.
Then, we refine our estimate using $\mathcal{M}(\check{\theta}_{\rm p})$ on the remaining $n-f(n)$ states.
The estimator $ \check{\theta}_{\rm r}\left( \check{\theta}_{\rm p} \right)$ employed in the refinement stage depends on the outcome of the preliminary estimator $ \check{\theta}_{\rm p} $. The outcome of $\check{\theta}_{\rm r}$ conditioned on $\check{\theta}_{\rm p}$ is described by the random variable $\check{\Theta}_{\rm r}\left(\check{\theta}_{\rm p}\right)$ with conditional density function $p_{\check{\Theta}_{\rm r}|\check{\Theta}_{\rm p}}\left(\check{\theta}_{\rm r}\mid
\check{\theta}_{\rm p}\right)$. Thus, we define MSE $ V_{\theta_{\mathrm{t}}}\left( \check{\theta}\right)  $ as: 
\begin{align}
V_{\theta_{\mathrm{t}}}\left(\check{\theta}_{\rm r}\right) 
    &= \int_{\Gamma}  V_{\theta_{\mathrm{t}}}\left( \check{\theta}_{\rm r}\left(\check{\theta}_{\rm p}\right)\right) p_{\check{\Theta}_{\rm p}}\left( \check{\theta}_{\rm p}\right) \dif \check{\theta}_{\rm p}, \label{eq: MSE definition}
\end{align}
where the MSE conditioned on the outcome of the preliminary estimator is:
\begin{align}
    V_{\theta_{\mathrm{t}}}\left( \check{\theta}_{\rm r}\left(\check{\theta}_{\rm p}\right)\right) = \int_{\Gamma} \left(\check{\theta}_{\rm r} - \theta_{\mathrm{t}}\right)^2 p_{\check{\Theta}_{\rm r}|\check{\Theta}_{\rm p}}( \check{\theta}_{\rm r}  \mid \check{\theta}_{\rm p}) \dif \check{\theta}_{\rm r}.
\end{align}
If the measurement also depends on nuisance parameters $\vec{\vartheta}$, we estimate these in the preliminary stage as well using $uf(n)$ additional states. In this case:
\begin{multline}
V_{\theta_{\mathrm{t}}}\left(\check{\theta}_{\rm r}\right) \\= \idotsint\limits_{\Gamma\times\Gamma_1\times\cdots\times\Gamma_u} V_{\theta_{\mathrm{t}}}\left( \check{\theta}_{\rm r}\left(\check{\theta}_{\rm p};\check{\vec{\vartheta}}\right)\right)p_{\check{\Theta}_{\rm p},\check{\vec{\varTheta}}}\left( \check{\theta}_{\rm p},\check{\vec{\vartheta}}\right) \dif \check{\theta}_{\rm p}\dif^u \check{\vec{\vartheta}},
    \label{eq: MSE definition nuisance}
\end{multline}
where the random variable $\check{\vec{\varTheta}}$ describes the outcome of nuisance parameter vector estimator $\check{\vec{\vartheta}}$,  $\dif^u\check{\vec{\vartheta}} \equiv \dif\vartheta_1\dots \dif\vartheta_u $, and
\begin{align}
    V_{\theta_{\mathrm{t}}}\left( \check{\theta}_{\rm r}\left(\check{\theta}_{\rm p};\check{\vec{\vartheta}}\right)\right) &= \int_{\Gamma} \left(\check{\theta}_{\rm r} - \theta_{\mathrm{t}}\right)^2  p_{\check{\Theta}_{\rm r}|\check{\Theta}_{\rm p},\check{\vec{\varTheta}}}( \check{\theta}_{\rm r}  \mid \check{\theta}_{\rm p},\check{\vec{\vartheta}}) \dif \check{\theta}_{\rm r}.
\end{align}
We conclude this section by defining
 balls around the true values of desired parameter $\theta$ and nuisance parameters $\vec{\vartheta}$ in their respective spaces:
\begin{align}
    \Gamma_{\delta}(\theta)&=\{\theta\in\Gamma:| \theta - \theta_{\mathrm{t}} |\leq\delta\} \label{eq:ball around theta0}\\
    \Gamma_{\delta}(\vartheta_i)&=\{\vartheta\in\Gamma_i:| \vartheta_i - \vartheta_{\mathrm{t},i} |\leq\delta\},\quad i = 1,\dots,u,\label{eq:ball around vartheta}
\end{align}
where  $\vartheta_{\mathrm{t},i}$ is the true value of $\vartheta_i$.

\subsection{Prior Work}
\label{sec:prior work}
To our knowledge, the convergence properties of the MSE $V_{\theta_{\mathrm{t}}}\left(\check{\theta}_{\rm r}\right)$ of the quantum two-stage estimator defined in Section \ref{sec:twostage} were first studied in detail by Hayashi and Matsumoto in \cite{hayashi2005statistical}.  We now restate their main result as a lemma. We adapt it to single-parameter estimation since this is the primary focus of our work.  We also make other changes, as discussed below.  

\begin{lemma}[\protect{\hspace{1sp}\cite[Th.~2]{hayashi2005statistical}}\label{lemma:HM}]
    The MSE of the two-stage estimator $\check{\theta}_{\rm r}$ satisfies:
\begin{align}
     \lim_{n\to\infty} n V_{\theta_{\mathrm{t}}}( \check{\theta}_{\rm r}) =  \mathcal{J}_{\theta_{\mathrm{t}}}^{-1} \label{eq:hayashi statement}
\end{align}
if the following conditions hold:
\begin{enumerate}
	   \item[HM1]\label{item:hayashi 1 pre} Preliminary estimator satisfies	$  	\lim_{n\to \infty} n\Pr\{| \check{\Theta}_{\rm p} - \theta_{\mathrm{t}}|>\epsilon_0 \}= 0, \ \ \forall \epsilon_0>0  $.
	\item[HM2]  \label{item:hayashi 2 para} MSE is bounded by a constant: $V_{\theta_{\mathrm{t}}}\left( \check{\theta}_{\rm r}\right) \le C_1, \forall \check{\theta}_{\rm r} \in \Gamma$.
	\item[HM3]\label{item:hayashi 4 continuous} Conditional MSE $V_{\theta_{\mathrm{t}}}\left(\check{\theta}_{\rm r}\left(\check{\theta}_{\rm p}\right)\right)$ is uniformly bounded: there exists $n_0>0$ s.t., for all $\delta_1,\epsilon_1>0$, $\check{\theta}_{\rm p}\in\Gamma_{\delta_{1}}(\theta)$,
 \begin{align*}
 \left| \left(n-f(n)\right)V_{\theta_{\mathrm{t}}}\left(\check{\theta}_{\rm r}\left(\check{\theta}_{\rm p}\right)\right) -  \mathcal{I}_{\theta_{\mathrm{t}},\check{\theta}_{\rm p}}^{-1}\right|&<\epsilon_{1}, \forall n > n_0.
 \end{align*}
    \item[HM4] $\mathcal{I}_{\theta_{\mathrm{t}},\check{\theta}_{\rm p}}$ is continuous over $\check{\theta}_{\rm p}$.
	\end{enumerate}
\end{lemma}

Before proving Lemma \ref{lemma:HM}, we contrast it with \cite[Th.~2]{hayashi2005statistical}.  
First, \cite{hayashi2005statistical} studies convergence of MSE to a multi-parameter \emph{quasi}-Cram\'{e}r-Rao bound \cite{nagaoka2005new, nagaoka2005parameter}. For a single parameter, this bound coincides with the standard results in \eqref{eq:qCRB}.  Thus, the right hand side (r.h.s.) of \eqref{eq:hayashi statement} is $\mathcal{J}_{\theta_{\mathrm{t}}}^{-1}$ and we omit the regularity condition B.5 in \cite{hayashi2005statistical}.
Instead, we add condition HM4, which is not onerous.
 Our condition HM1 is the condition B.1 in \cite{hayashi2005statistical} with factor $n$ included in front of probability (this is a typo in \cite{hayashi2005statistical}, as the proof of \cite[Th.~2]{hayashi2005statistical}  in \cite[Sec.~3.4]{hayashi2005statistical} does not hold without it).
   Condition HM2 relaxes condition B.2 in \cite{hayashi2005statistical}; the proof of \cite[Th.~2]{hayashi2005statistical} holds with this relaxation. 
Condition B.3 in \cite{hayashi2005statistical} is omitted since it is not used in the proof of \cite[Th.~2]{hayashi2005statistical}.  Our condition HM3 is condition B.4 in \cite{hayashi2005statistical} generalized to allow $f(n)\in\omega(1)\cap o(n)$ states to be used in the preliminary stage.  The authors of \cite{hayashi2005statistical} set $f(n)=\sqrt{n}$, although the proof of \cite[Th.~2]{hayashi2005statistical} holds for any $f(n)\in \omega(1)\cap o(n)$.

\begin{proof} We begin with achievability. Using \eqref{eq: MSE definition},
\begin{align}
    &\lim_{n\to\infty} n V_{\theta_{\mathrm{t}}}( \check{\theta}_{\rm r})\nonumber \\ 
   &= \lim_{n\to\infty} n \Bigg[ \int_{\Gamma_{\delta_1}^c(\theta)}  V_{\theta_{\mathrm{t}}}\left(\check{\theta}_{\rm r}\left(\check{\theta}_{\rm p}\right)\right) p_{\check{\Theta}_{\rm p}}( \check{\theta}_{\rm p}) \dif \check{\theta}_{\rm p} \nonumber\\
   &\phantom{ \lim_{n\to\infty} +++} +\int_{\Gamma_{\delta_1}(\theta)}  V_{\theta_{\mathrm{t}}}\left( \check{\theta}_{\rm r}\left( \check{\theta}_{\rm p}\right)\right) p_{\check{\Theta}_{\rm p}}( \check{\theta}_{\rm p}) \dif \check{\theta}_{\rm p} \Bigg], \label{eq:split}
\end{align}
where $\Gamma_{\delta}^c(\theta)$ is the complement of the ball $\Gamma_{\delta}(\theta)$ defined in \eqref{eq:ball around theta0}.
Consider the first limit in \eqref{eq:split}:
\begin{align}
    &\lim_{n\to\infty}n \int_{\Gamma_{\delta_1}^c(\theta)}  V_{\theta_{\mathrm{t}}}\left( \check{\theta}_{\rm r}\left( \check{\theta}_{\rm p}\right)\right) p_{\check{\Theta}_{\rm p}}( \check{\theta}_{\rm p}) \dif \check{\theta}_{\rm p} \nonumber\\
      &\le \lim_{n\to\infty}n C_1 \int_{\Gamma_{\delta_1}^c(\theta)}  p_{\check{\Theta}_{\rm p}} ( \check{\theta}_{\rm p}) \dif \check{\theta}_{\rm p} \label{eq:First limit u b}\\
     &= \lim\limits_{n\to \infty} nC_1\Pr\{| \check{\Theta}_{\rm p} - \theta_{\mathrm{t}}|>\delta_1 \}= 0, \label{eq: hayashi first term}
\end{align}
where \eqref{eq:First limit u b} and \eqref{eq: hayashi first term} are due to conditions HM2 and HM1, respectively.
Consider the second limit in \eqref{eq:split}:
\begin{align}
&\lim_{n\to\infty} n\int_{\Gamma_{\delta_1}(\theta)}  V_{\theta_{\mathrm{t}}}\left( \check{\theta}_{\rm r}\left( \check{\theta}_{\rm p}\right)\right) p_{\check{\Theta}_{\rm p}}( \check{\theta}_{\rm p}) \dif \check{\theta}_{\rm p} \nonumber\\
 &\le \lim_{n\to\infty} \frac{n}{n - f(n)} \int_{\Gamma_{\delta_1}(\theta)} (  \mathcal{I}_{\theta_{\mathrm{t}},\check{\theta}_{\rm p}}^{-1} + \epsilon_1 )  p_{\check{\Theta}_{\rm p}}(\check{\theta}_{\rm p}) \dif \check{\theta}_{\rm p} \label{eq:hayashi s u p}  \\
  & =  \! \lim_{n\to\infty} \frac{n  }{n - f(n)} (\mathcal{I}_{\theta_{\mathrm{t}},\theta_{\mathrm{t}}}^{-1}  \! + \epsilon_1 \!+ \epsilon_2 )\Pr\{\check{\Theta}_{\rm p}\in\Gamma_{\delta_1}(\theta) \} \label{eq:hayashi continuity} \\
  &= \mathcal{I}_{\theta_{\mathrm{t}},\theta_{\mathrm{t}}}^{-1}= \mathcal{J}_{\theta_{\mathrm{t}}}^{-1}  \label{eq:hayashi second term},
\end{align}
where \eqref{eq:hayashi s u p} and \eqref{eq:hayashi continuity} are due to conditions HM3 and HM4, with $\epsilon_2>0$ arbitrarily small.
Substitution of \eqref{eq: hayashi first term} and \eqref{eq:hayashi second term} into \eqref{eq:split} yields the achievability of
$    \lim_{n\to\infty} n V_{\theta_{\mathrm{t}}}( \check{\theta}_{\rm r}) \le \mathcal{J}_{\theta_{\mathrm{t}}}^{-1}$. The QCRB in \eqref{eq:qCRB} yields the converse and the theorem.
\end{proof}

\section{Asymptotic Consistency and Normality of Two-Stage Quantum Estimator}
\label{sec:main}
Numerical evidence suggests that Lemma \ref{lemma:HM} holds for certain quantum estimation problems (e.g., transmittance sensing, see \cite[Fig.~10]{gong22losssensing}). However, its stringent conditions pose significant barriers to its use. First, condition HM1 is stricter than the standard asymptotic consistency. More importantly, uniform integrability of the estimator $\check{\theta}_{\rm r}$ used in the refinement stage is necessary for condition HM3 to hold. 
Indeed, although the authors of \cite{hayashi2005statistical} suggest using an MLE in \cite[Sec.~3.2]{hayashi2005statistical}, they recognize that their condition B.4 (our condition HM3) is difficult to verify. It is well-known that, although MLE is asymptotically consistent, typically, it does not satisfy condition HM3 (for instance, see remarks after \cite[Prop.~IV.D.2]{poor94intro-det-est}).

At the same time, asymptotic consistency and normality of an estimator are sufficient in many practical settings (e.g., to approximate confidence intervals \cite[Sec.~1]{newey94largesampleestimation}).  Focusing on these allows us to relax the conditions of Lemma \ref{lemma:HM} and to include the analysis of nuisance parameters (which are also estimated in the first stage).  In fact, under certain regularity conditions, MLEs are asymptotically consistent and normal. Thus, when used on the outcomes of the SLD-eigendecomposition quantum measurement, the following allows us to claim quantum optimality with a suitable preliminary estimator. 
We denote by $X_n\xrightarrow{a.s.} X$, $X_n\xrightarrow{p} X$, and $X_n\xrightarrow{d} X$ convergence of a sequence of random variables $(X_n)$ to $X$ almost surely (a.s.), in probability, and in distribution, respectively \cite{billingsley95measure}.  We also denote a Gaussian (normal) distribution with mean $\mu$ and variance $\sigma^2$ by $\mathcal{N}(\mu,\sigma^2)$.

\begin{theorem} \label{theorem: our}
 The outcome of the refinement stage in the two-stage quantum estimator is weakly (strongly) consistent and asymptotically normal:
 \begin{align}
 \check{\Theta}_{\rm r} &\xrightarrow{p~(a.s.)} \theta_{\mathrm{t}} \label{eq:asymptotic consistency}\\
 \sqrt{n - f(n)} \left( \check{\Theta}_{\rm r} - \theta_{\mathrm{t}} \right) &\overset{d}{\to} \mathcal{N}\left(0, \mathcal{J}_{\theta_{\mathrm{t}}}^{-1}\right) \label{eq:asymptotic normality}
 \end{align}
 for $f(n)\in\omega(1)\cap o(n)$, if the following conditions hold:
 \begin{enumerate}
  \item \label{item: 1 pre} The preliminary estimator is weakly (strongly) consistent: $  \check{\Theta}_{\rm p} \xrightarrow{p~(a.s.)}  \theta_{\mathrm{t}}. $
    \item \label{item: 3 nuisance} The vector estimators $ \check{\vec{\varTheta}} = \{\check{\varTheta}_1,\dots,\check{\varTheta}_u\} $ of the nuisance parameters $\vec{\vartheta}  = \{ \vartheta_i\}_{i=1}^u $ are weakly (strongly) consistent: $  \check{\vartheta}_i \xrightarrow{p~(a.s.)} \vartheta_{\mathrm{t},i} $ for all $i$, where $ \vartheta_{\mathrm{t},i}$ is $\vartheta_i$'s true value.
  \item \label{item: 2 MLE} There exists $ \delta_2>0 $ such that, when the preliminary estimate is close to $ \theta_{\mathrm{t}} $ and $\check{\vartheta}_i$ are close to $ \vartheta_{\mathrm{t},i}$ for all $i$, i.e., $ \check{\theta}_{\rm p}\in\Gamma_{\delta_2}(\theta)$ and $\check{\vartheta}_{i}\in\Gamma_{\delta_2}(\vartheta_i) $, the refinement estimator $ \check{\Theta}_{\rm r}\left( \check{\theta}_{\rm p},\check{\vec{\vartheta}} \right) $ has the following properties: 
  \begin{enumerate}
   \item \label{condition:consistency} Weak (strong) consistency:
            \begin{align*}
            \check{\Theta}_{\rm r}\left( \check{\theta}_{\rm p},\check{\vec{\vartheta}}\right)&\xrightarrow{p~(a.s.)}  \theta_{\mathrm{t}}.
            \end{align*}
   \item \label{condition:asymptotic normality}  Asymptotic normality:
    \begin{align*} \sqrt{n -f(n)} \left( \check{\Theta}_{\rm r}\left(\check{\theta}_{\rm p},\check{\vec{\vartheta}}\right) - \theta_{\mathrm{t}} \right) &\xrightarrow{d} \mathcal{N}\left(0, \mathcal{I}_{\theta_{\mathrm{t}},\check{\theta}_{p},\check{\vec{\vartheta}}}^{-1}\right),
             \end{align*}
             where $ \mathcal{I}_{\theta_{\mathrm{t}},\check{\theta}_{\rm p},\check{\vec{\vartheta}}} $ is the classical FI associated with $\theta$ for a random variable describing the outcome of $ \mathcal{M}_{\rm r}(\check{\theta}_{\rm p}, \check{\vec{\vartheta}})$.
  \end{enumerate} 
  \item \label{item:3 continuous} $ \mathcal{I}_{\theta_{\mathrm{t}},\check{\theta}_{p},\check{\vec{\vartheta}}} $ is continuous over $ \check{\theta}_{\rm p} $ and $ \check{\vartheta}_i $ for all $i$.
 \end{enumerate}  
\end{theorem}

The proof of Theorem \ref{theorem: our} is in Appendix \ref{app: main proof}.
Next, we apply it to obtain the asymptotics of quantum-enhanced transmittance sensing in a lossy thermal-noise bosonic channel with an unknown phase offset, a problem of significant practical importance that we explored in \cite{gong21losssensing,gong22losssensing}.

\section{Asymptotics of Quantum-enhanced Transmittance Sensing}
\label{sec:transmittance sensing}
\begin{figure}[h]
 \centering
    \includegraphics[width=1.0\columnwidth]{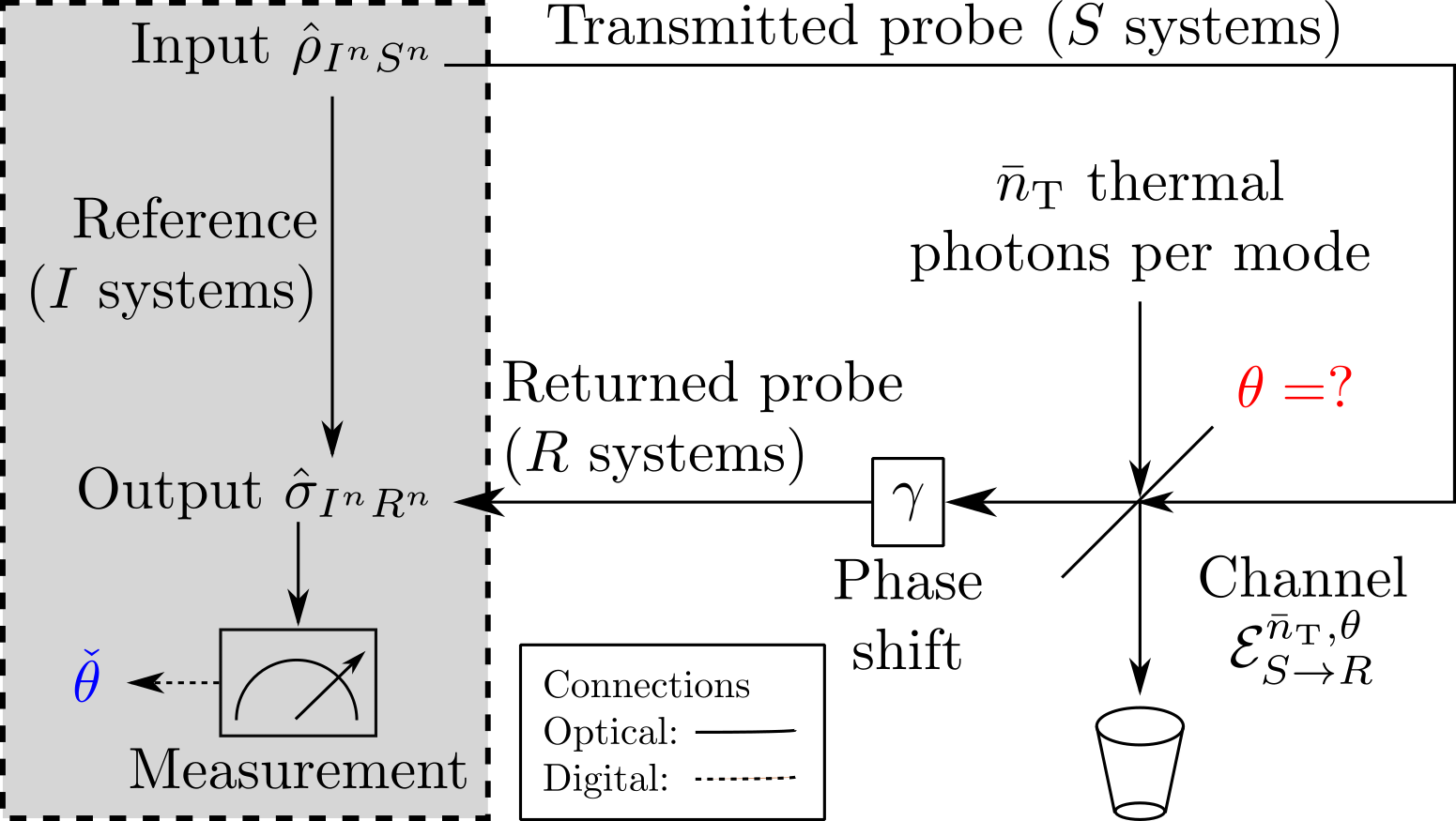}
 \caption{\justifying Sensing of unknown transmittance $\theta$. Sensor transmits $n$-mode probes (systems $S$ of bipartite state $\hat{\rho}_{I^nS^n}$) into a  lossy thermal-noise bosonic channel $\mathcal{E}^{(\bar{n}_{\rm T},\theta)}$ modeled by a beamsplitter with unknown transmittance $ \theta $ mixing signal and a thermal state with mean thermal photon
  number $\bar{n}_{\rm T}\equiv\frac{\bar{n}_{\rm B}}{1-\theta}$. Additionally, the returned probe undergoes an unknown phase shift $\gamma$. Reference idler systems $I$ are used in the measurement of output state $\hat{\sigma}_{I^nR^n}(\theta)$, with outcomes passed to estimator $\check{\theta}$. Solid and dashed lines denote optical (quantum) and digital (classical) connections, respectively.
  \label{fig:setup}}
 \vspace{-0.1cm}
\end{figure}

Consider the system model in Fig.~\ref{fig:setup}.
As in \cite{gong21losssensing, gong22losssensing}, we employ
a lossy thermal-noise  bosonic channel $\mathcal{E}^{(\bar{n}_{\rm T},\theta)}$, a quantum-mechanical description of many practical channels, including optical, microwave, and radio-frequency.
A tutorial overview of the bosonic channels is available in \cite{weedbrook12gaussianQIrmp} while \cite{serafini2023quantum} provides their rigorous exploration.
We wish to estimate the unknown power transmittance $\theta$ of this channel.
To prevent the noise from carrying useful information about $\theta$ to the sensor (so-called ``shadow effect'' \cite{jonsson22gaussianlosssensing}), we set the thermal environment mean photon number $\bar{n}_{\rm T}\equiv\frac{\bar{n}_{\rm B}}{1-\theta}$, as in the quantum illumination literature \cite{nair20qi, lloyd08quantumillumination, tan08qigaussianstates, guha09quantumilluminationOPA, shapiro20QIstory, sanz17estimationQI}.
In this paper, we add an unknown phase shift $\gamma$ to the bosonic channel model, which arises frequently in practice, e.g., due to incomplete knowledge of the distance between the sensor and the target.
Here, $\gamma$ is a nuisance parameter that must be estimated in the first stage along with the parameter of interest, the transmittance $\theta$.
In fact, the need to address this problem inspired us to include the nuisance parameters in Theorem \ref{theorem: our}.

\begin{figure}[h]
 \centering
 \includegraphics[width=1\columnwidth]{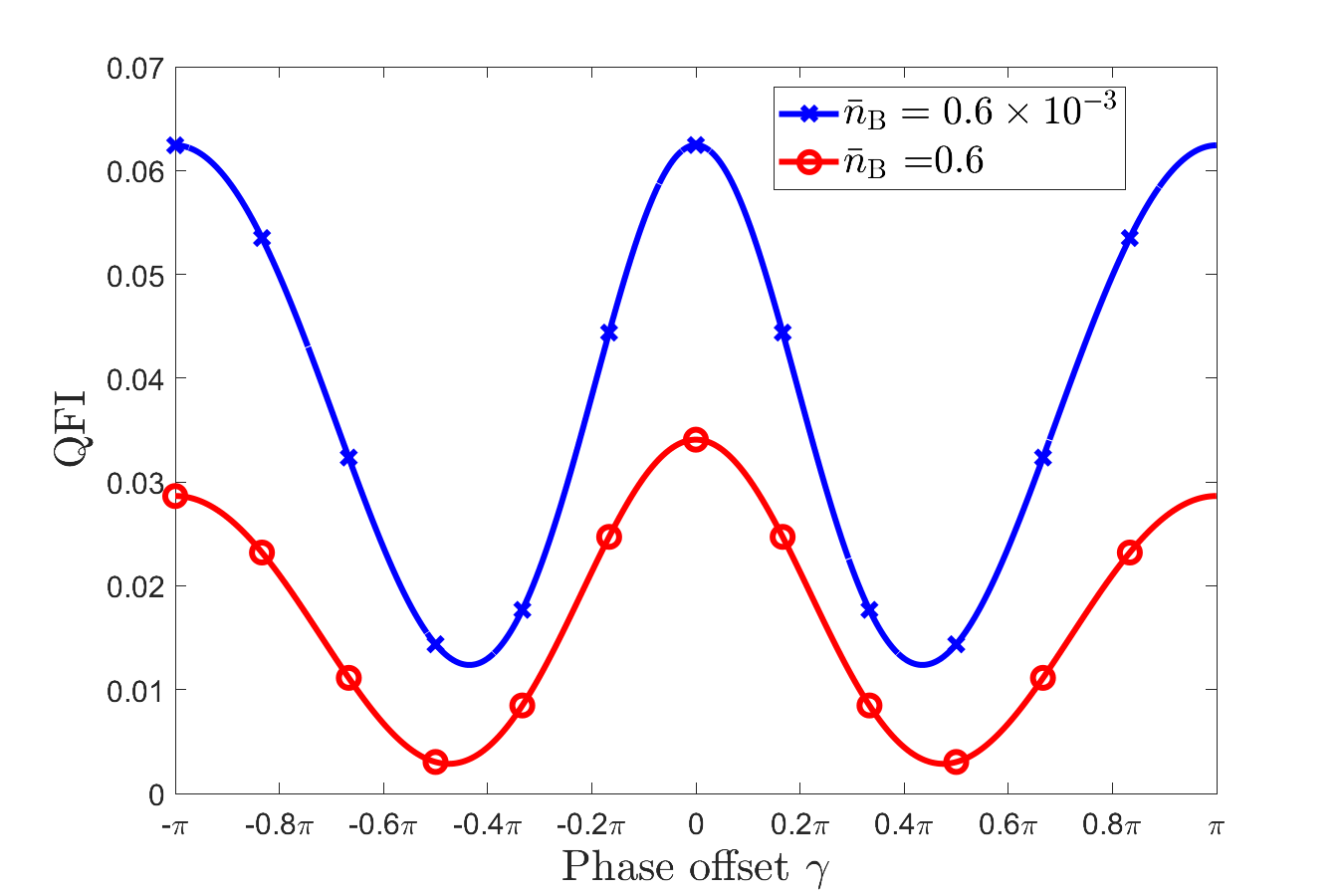}
 \caption{Illustration of the impact of unknown phase shift $\gamma$ on the QFI associated with transmittance. We use a two-mode squeezed vacuum (TMSV) probe with mean $\bar{n}_{\rm S}=10^{-2}$ photons per mode to estimate transmittance $\theta$ of a lossy thermal-noise bosonic channel. We set $\theta=0.2$ and the mean thermal background photon number $\bar{n}_{\rm B}$ to different values.
    \label{fig:QFI}}
\end{figure}

\begin{figure*}[t]
\centering
 \subfloat[Refinement stage]{\includegraphics[width=0.95\columnwidth]{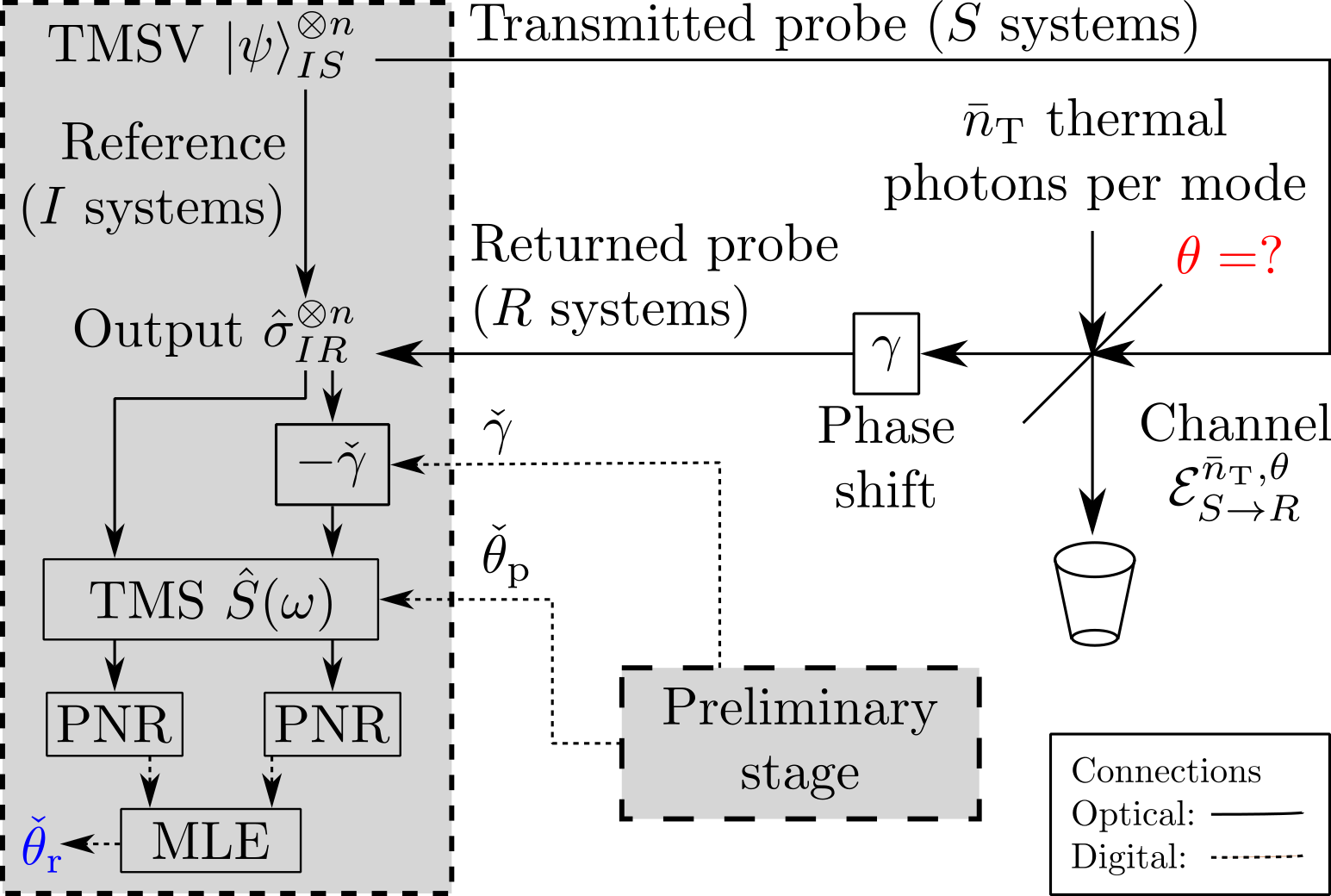}\label{subfig:optimal}}\hfill
 \subfloat[Preliminary stage]{\includegraphics[width=0.95\columnwidth]{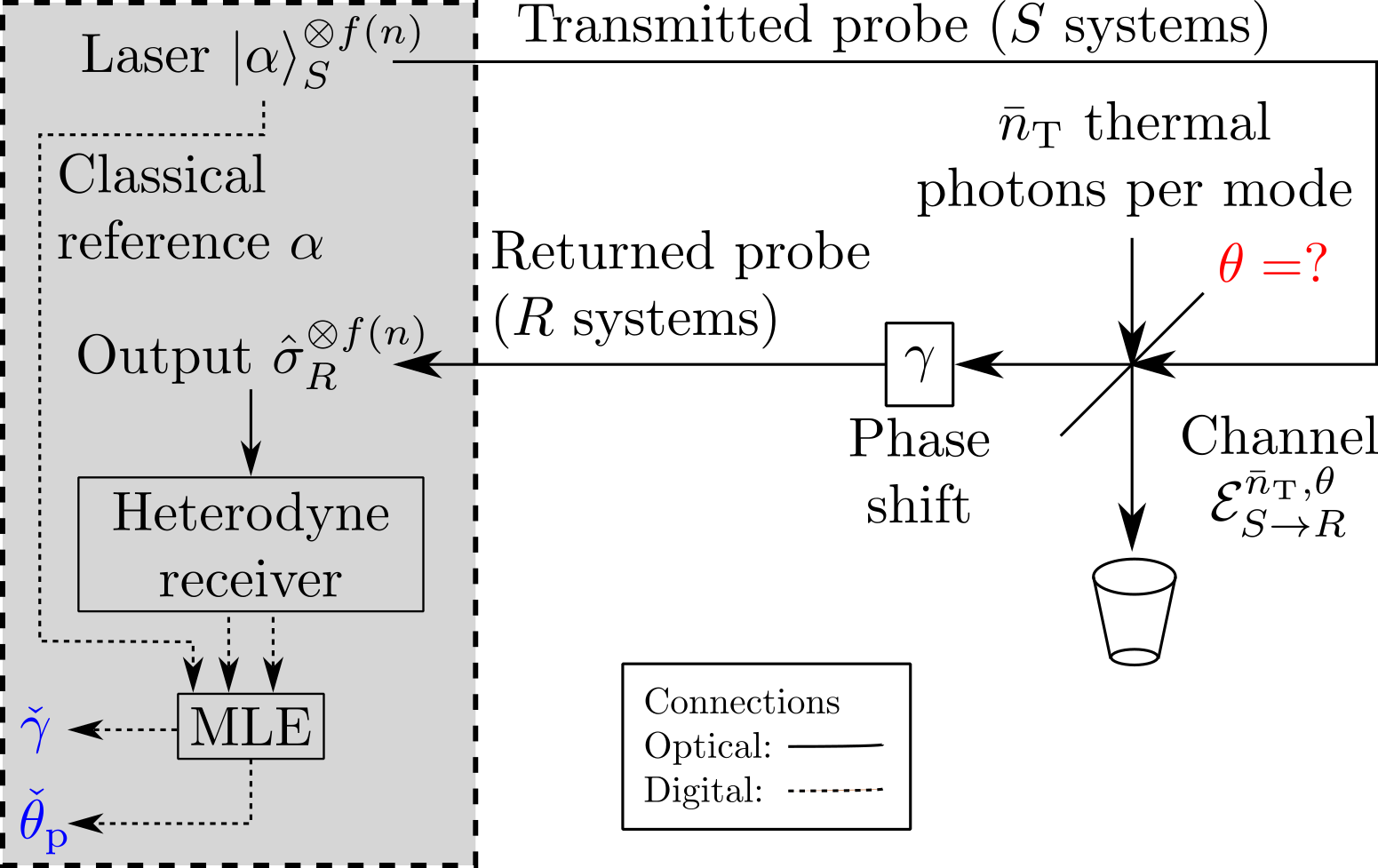}\label{subfig:preliminary}}   
 \caption{\justifying Two-stage transmittance sensor.  Refinement stage in \protect\subref{subfig:optimal} uses $(\check{\gamma},\check{\theta}_{\rm p})$ from the preliminary stage in \protect\subref{subfig:preliminary} and $n-f(n)$ two-mode squeezed vacuum (TMSV) states $\ket{\psi}_{IS}$ to estimate the unknown transmittance $\theta$, where $f(n)\in\omega(1)\cap o(n)$.  When $
\ket{\psi}_{IS}$ is transmitted, a bipartite output state $\hat{\sigma}_{IR}$ occupies a retained idler system $I$ and a corresponding returned probe system $R$.  The receiver applies correction in $R$ for the channel phase shift $\gamma$ using the estimate $\check{\gamma}$. It then applies a two-mode squeezer (TMS) separately to each of the $n$ output states $\hat{\sigma}_{IR}$ with parameter $\omega$ that is calculated using the preliminary estimate $\check{\theta}_{\rm p}$, followed by independent photon-number-resolving (PNR) measurement of each output mode.  MLE is used on the resulting classical output to obtain the estimate $\check{\theta}_{\rm r}$. 
Preliminary stage in \protect\subref{subfig:preliminary} uses $f(n)\in\omega(1)\cap o(n)$ coherent states $\ket{\alpha}_{S}$ and a heterodyne measurement to estimate the unknown transmittance $\theta$ and phase shift $\gamma$.  The output state $\hat{\sigma}_{R}$ is a displaced thermal state. A pair of Gaussian random variables describes the output of heterodyne measurement \cite{weedbrook12gaussianQIrmp},\cite[Ch.~7.3.2]{guha04mastersthesis}.  An MLE $\check{\theta}_{\rm p}$ that uses the heterodyne receiver's output and the value of $\alpha$ as a classical reference is analyzed in Appendix \ref{app:pre consistency}.
As in Fig.~\ref{fig:setup}, $\bar{n}_{\rm T}\equiv\frac{\bar{n}_{\rm B}}{1-\theta}$, and solid and dashed lines denote optical (quantum) and digital (classical) connections, respectively.   \label{fig:receiver}}
\end{figure*}

The sensor uses a bipartite quantum state $\hat{\rho}_{I^nS^n}$ occupying $n$ signal and idler systems $S$ and $I$.
Signal systems $S$ interrogate the target using $n$ available modes of channel $\mathcal{E}^{\bar{n}_{\rm T},\theta}_{S\rightarrow R}$, while
the idler systems $I$ are retained as a lossless and noiseless reference.
An unknown phase shift $\gamma$ is an offset between signal and idler modes.
If $\gamma$ is known, an optical phase shifter is used to reset this offset to zero.
When $\gamma=0$, the optimal quantum state for transmittance sensing with small mean probe photon number per mode $\bar{n}_{\rm S}\to 0$ is the two-mode squeezed vacuum (TMSV) state \cite{gong21losssensing} (it is an optimal Gaussian quantum state for any $\bar{n}_{\rm S}$ \cite{jonsson22gaussianlosssensing} ).
Fig.~\ref{fig:QFI} illustrates the deleterious impact unknown $\gamma$ has on TMSV-based sensors; we defer the detailed study to \cite[Ch.~5]{gong_thesis}.
For $\gamma=0$, the corresponding optimal POVM from the eigen-states of the SLD, per Section \ref{sec:quantum estimation prerequisites}, is a two-mode squeezer with squeezing parameter $\omega$ followed by the photon-number-resolving (PNR) measurements of each output mode \cite{gong22losssensing}.
Fig.~\ref{fig:receiver}\subref{subfig:optimal} diagrams this sensor.

Our measurement consists of well-known optical components.
In fact, the setup in Fig.~\ref{fig:receiver}\subref{subfig:optimal} is a form of an SU(1,1) interferometer.
Its SU(2) counterpart replaces the two-mode squeezers with beam-splitters.
SU(1,1) and SU(2) interferometers \cite{yurke86SU2SU22interferometers} are used in quantum metrology \cite{giovannetti11metrology} to measure phase $\gamma$.
Transmittance $\theta$ is often the nuisance parameter in this setting, and active squeezing in SU(1,1) interferometer allows mitigation of photon losses \cite{sparaciari16gaussianinterferometry, Manceau_2017}.
The reversal of the typical roles of transmittance $\theta$ and phase $\gamma$ in \cite{gong21losssensing, gong22losssensing}, and here, shows the versatility of SU(1,1) interferometry.
Although experimental demonstrations of SU(1,1) interferometers in various forms \cite{Manceau_2017, Hudelist2014, Chen15, Linnemann16} strongly suggest feasible practical implementation of QCRB-achieving transmittance sensors, there are three caveats:
\begin{enumerate}
\item Whether the measurement in Fig.~\ref{fig:receiver}\subref{subfig:optimal} exists (i.e., whether a solution for $ \omega $ can be found) depends on the values of $ \theta $, $ \bar{n}_{\rm S} $ and $ \bar{n}_{\rm B} $, as illustrated in \cite[Fig.~4]{gong22losssensing}. Thus, the parameter space that this measurement covers is $ \Gamma = [\theta^{\prime\prime},1]$, where $ \theta^{\prime\prime}>0$ is a function of $\bar{n}_{\rm S}$ and $\bar{n}_{\rm B} $. 
A different, possibly suboptimal, measurement must be used outside this parameter space. We explore this in \cite[Ch.~5]{gong_thesis}.
\item The measurement structure determined by $ \omega $ depends on the parameter of interest $ \theta $. 
\item The measurement is optimal only for $\gamma=0$.  Thus, we must use an optical phase shifter on either the signal or the idler mode to reverse the phase offset (in Fig.~\ref{fig:receiver}\subref{subfig:optimal}, we choose the signal mode). Although it is a well-known optical component, an estimate for $\gamma$ is required. 
\end{enumerate}
The two-stage method introduced in Section \ref{sec:twostage} addresses the last two caveats.

In the preliminary stage, shown on Fig.~\ref{fig:receiver}\subref{subfig:preliminary}, we employ a laser-light (coherent-state) probe with mean photon number per mode $\bar{n}_{\rm S}$. 
 We prove that the MLEs for both $\theta$ and $\gamma$ applied to heterodyne measurement outcomes are strongly consistent in Appendix \ref{app:pre consistency}, thus satisfying the strong versions of conditions \ref{item: 1 pre} and \ref{item: 3 nuisance} of Theorem \ref{theorem: our}. 
We use our optimal measurement in the refinement stage.
The statistics of its output are described by an i.i.d.~sequence of data pairs corresponding to the photon counts in each PNR detector $\left\{\left(K_i,M_i\right)\right\}_{i=f(n)+1}^n$ (although $K_i$ and $M_i$ in each pair are correlated).  
We derive the p.m.f.~in Appendix \ref{app:pmf}: 
\begin{align}
    & p_{K,M}(k,m;\theta,\check{\theta}_{\rm p}, \check{\gamma}) \nonumber\\
 &=\bra{k,m}\hat{S}_{IR}^{\dagger}(\omega)\hat{U}_R\left(-\check{\gamma}\right)\hat{\sigma}_{IR}\hat{U}_R\left(\check{\gamma}\right)\hat{S}_{IR}(\omega) \ket{k,m}   \\
 &=\sum_{s=0}^{\infty} r_{s,t^{\prime}}|\bra{s,t^{\prime}}\hat{S}_{IR}(\zeta)\hat{U}_R(\check{\gamma}-\gamma_{\rm t})\hat{S}_{IR}(\omega)\ket{k,m} |^2, \label{eq:likelihood function}
\end{align}
where 
\begin{align}
r_{s,t} &= \frac{N_1^s N_2^t}{(1+N_1)^{s+1}(1+N_2)^{t+1}},\label{eq:rst}
\end{align}
with $N_1$ and $N_2$ functions of $\theta_{\rm t}$, $\bar{n}_{\rm S}$, and $\bar{n}_{\rm B}$ given in \cite[Eqs.~(59) and (60)]{gong22losssensing}.
Furthermore, $\hat{S}_{IR}(\cdot)$ is a two-mode squeezing operator acting on idler and returned probe systems $I$ and $R$, $\hat{U}_R(\cdot)$ is a phase shift operator acting on returned probe system $R$ only (formally, one would write $\hat{I}_I\otimes\hat{U}_R(\cdot)$ instead of $\hat{U}_R(\cdot)$, however, we drop the identity operator $\hat{I}_I$ acting on the idler system $I$ for brevity),  $\omega$ is a function of $\check{\theta}_{\mathrm{p}}$, $\bar{n}_{\rm S}$, and $\bar{n}_{\rm B}$ described in \cite[Sec.~IV.A]{gong22losssensing}, $\zeta$ is a function of $\theta_{\rm t}$, $\bar{n}_{\rm S}$, and $\bar{n}_{\rm B}$ given in \cite[Eq.~(56)]{gong22losssensing}, $\gamma_{\rm t}$ is the true value of phase shift $\gamma$, $t^{\prime} = s-k+m$,
\begin{widetext}
\begin{align}
\bra{s,t}\hat{S}_{IR}(\zeta)\hat{U}_R(\check{\gamma}-\gamma_{\rm t})\hat{S}_{IR}(\omega)\ket{k,m}
&=\sum_{i_1=0}^{\min(k,m)}  \sum_{a_1=0}^{\infty} \sum_{i_2=\max(0,k-s-i_1+a_1)}^{\min(k-i_1+a_1,m-i_1+a_1)}  \frac{(-\tau_1)^{a_1}(\tau_1^*)^{i_1}}{a_1!i_1!\nu_1^{k + m - 2i_1 + 1}}  \nonumber      \\
&\phantom{=}\times   \frac{(-\tau_2)^{s-k+i_1-a_1+i_2}(\tau_2^*)^{i_2}}{(s-k+i_1-a_1+i_2)!i_2!\nu_2^{k + m -2i_1 +2 a_1 - 2i_2 + 1}} \nonumber \\
&\phantom{=}\times \frac{\sqrt{k!m!(k-i_1+a_1)!(m-i_1+a_1)!(k-i_1+a_1)!(m-i_1+a_1)!s!t!}}{(k-i_1)!(m-i_1)!(k-i_1+a_1-i_2)!(m-i_1+a_1-i_2)!e^{j(\check{\gamma}-\gamma_{\rm t})(k-i_1+a_1)}} \label{eq:summand},
\end{align}
\end{widetext}
$ \tau_1 = \tanh\omega$, $ \tau_2 = \tanh\zeta$, $\nu_1 = \cosh\omega$, and $\nu_2 = \cosh \zeta$.
Note that our two stages differ not only in the measurement structure but also in the quantum state being measured.
Although the MLE of $\theta$ for the optimal receiver has no closed form, we prove its strong consistency and asymptotic normality in Appendices \ref{app:consistency} and \ref{app:normality proof}, meeting condition \ref{item: 2 MLE} of Theorem \ref{theorem: our}. Furthermore, the squeezing parameter $\omega$ is continuous in $\check{\gamma}$ and the preliminary estimate $\check{\theta}_{\rm p}$, and the classical FI associated with $\theta$ in $(K,M)$ is continuous in $\omega$, satisfying condition \ref{item:3 continuous} of Theorem \ref{theorem: our}. Thus, our two-stage transmittance estimator is strongly consistent, asymptotically normal, and, hence, quantum-optimal in the sense of Theorem \ref{theorem: our}.
Indeed, this matches the numerical results in \cite[Fig.~10]{gong22losssensing}.

\section{Conclusion} \label{sec: conclusion}
Quantum estimation theory yields optimal measurements of a scalar parameter embedded in a quantum state \cite{helstrom76quantumdetect}.
However, often, these measurements depend on the parameter of interest.
This necessitates a two-stage approach \cite{gill00twostagemeasurement, hayashi2005statistical}, \cite[Ch.~6.4]{hayashi17qit}, where a preliminary estimate is derived from a sub-optimal measurement, and is then used to construct an optimal measurement that yields a refined estimate.
Here, we establish the conditions for the strong and weak consistency as well as the asymptotic normality of this two-stage approach, with QCRB being the variance of the limiting Gaussian in the latter claim.
This matches the usual asymptotic properties of the MLEs.
We include estimation of nuisance parameters in the first stage and apply our methodology to show that the quantum-enhanced transmittance estimator from \cite{gong22losssensing} is strongly consistent and asymptotically normal, even when an unknown phase shift is introduced by the bosonic channel.
Our work also immediately strengthens existing results such as \cite{grace20adaptivesuperresolution, Sajjad21, Sajjad24, ozer2024adaptivesuperresolutionimagingprior, deshler2025quantumlimitedspatialresolution} that employ two-stage approaches to achieve QCRB, by establishing a framework for their asymptotic analysis.

Although out of scope in this paper, extending our results to multiple parameters is an intriguing area for future work.
Attaining multi-parameter QCRB \cite[Ch.~VIII.4(a)]{helstrom76quantumdetect} is complicated by the non-commutativity of quantum measurements for each parameter.
A potential direction of research would focus on the asymptotics of quantum estimators in the context of Holevo-Cram\'{e}r-Rao \cite{holevo82quantumstat} and quasi-Cram\'{e}r-Rao \cite{nagaoka2005new, nagaoka2005parameter} bounds.
In the immediate term, our results will allow establishing optimality claims for various single-parameter quantum estimation problems.
Indeed, we will apply them to robust quantum-inspired super-resolution imaging \cite{tan22robustspade-asilomar}.

\section*{Acknowledgments}
The authors are grateful to Brian J.~Smith and Michael G.~Raymer for discussions during BAB's visit to the University of Oregon in January 2024, which inspired the inclusion of phase shift in the model in Section \ref{sec:transmittance sensing} and the subsequent treatment of the nuisance parameters in Theorem \ref{theorem: our}.
The authors also acknowledge helpful discussions with Nathaniel Rodriguez, Christos N.~Gagatsos, Amit Ashok, Michael R.~Grace, Michael S.~Bullock, and Saikat Guha.
This work was supported by the National Science Foundation under Grant No.~CCF-2045530.

 \appendix

\section{Proof of Asymptotic Consistency and Normality of Two-Stage Quantum Estimator}
\label{app: main proof}
Note: we prove the weak (strong) consistency $\check{\Theta}_{\rm r} \xrightarrow{p~(a.s.)}\theta_{\mathrm{t}} $ using the in-probability (almost-sure) versions of conditions \ref{item: 1 pre}, \ref{item: 3 nuisance},  and \ref{condition:consistency}.
\begin{proof}[Proof (Theorem \ref{theorem: our}).] 
Let $\Omega_{\delta} = \cap_{i=1}^u \Gamma_{\delta}(\vartheta_i)\cap\Gamma_{\delta}(\theta)$, and $\Omega_{\delta}^c = \cup_{i=0}^u \Gamma_{\delta}^c(\vartheta_i)\cup\Gamma_{\delta}^c(\theta) $. 
First, we show the weak consistency of $ \check{\Theta}_{\rm r}$:
\begin{widetext}
\begin{align}
\Pr\left\{ \left|\check{\Theta}_{\rm r} - \theta_{\mathrm{t}} \right|>\epsilon_3 \right\}    &= \idotsint\limits_{\Gamma\times\Gamma_1\times\dots\times\Gamma_u} p_{\check{\Theta}_{\rm p}, \check{\vec{\varTheta}}}\left(\check{\theta}_{\rm p},\check{\vec{\vartheta}} \right)\Pr\left\{ \left|\check{\Theta}_{\rm r}\left( \check{\theta}_{\rm p};\check{\vec{\vartheta}} \right) - \theta_{\mathrm{t}} \right|>\epsilon_3 \right\} \dif \check{\theta}_{\rm p} \dif^u\check{\vec{\vartheta}} \nonumber \\
	&=  \idotsint_{\Omega_{\delta_2}^c} p_{\check{\Theta}_{\rm p}, \check{\vec{\varTheta}}}\left(\check{\theta}_{\rm p},\check{\vec{\vartheta}}  \right) \Pr\left\{ \left|\check{\Theta}_{\rm r}\left( \check{\theta}_{\rm p};\check{\vec{\vartheta}} \right) - \theta_{\mathrm{t}} \right|>\epsilon_3 \right\} \dif \check{\theta}_{\rm p}  \dif^u\check{\vec{\vartheta}}\nonumber \\
	&\phantom{=}+\idotsint_{\Omega_{\delta_2}} p_{\check{\Theta}_{\rm p},\check{\vec{\varTheta}}}\left(\check{\theta}_{\rm p},\check{\vec{\vartheta}} \right)\Pr\left\{ \left|\check{\Theta}_{\rm r}\left( \check{\theta}_{\rm p};\check{\vec{\vartheta}} \right) - \theta_{\mathrm{t}} \right|>\epsilon_3 \right\} \dif \check{\theta}_{\rm p} \dif^u\check{\vec{\vartheta}}, \label{eq:consistency 2}
\end{align}
where 
$ \delta_2 $ is from condition \ref{item: 2 MLE}. The limit of the first term in \eqref{eq:consistency 2} is:
\begin{align}
&\lim_{n\to\infty} \idotsint_{\Omega_{\delta_2}^c} p_{\check{\Theta}_{\rm p}, \check{\vec{\varTheta}}}\left(\check{\theta}_{\rm p},\check{\vec{\vartheta}} \right) 
\Pr\left\{ \left|\check{\Theta}_{\rm r}\left( \check{\theta}_{\rm p};\check{\vec{\vartheta}} \right) - \theta_{\mathrm{t}} \right|>\epsilon_3 \right\} \dif \check{\theta}_{\rm p} \dif^u\check{\vec{\vartheta}}\nonumber\\
&\phantom{===}\le \lim_{n\to\infty} \idotsint_{\Omega_{\delta_2}^c} p_{\check{\Theta}_{\rm p}, \check{\vec{\varTheta}}}\left(\check{\theta}_{\rm p},\check{\vec{\vartheta}} \right)  \dif \check{\theta}_{\rm p} \dif^u\check{\vec{\vartheta}}=  \lim_{n\to\infty} \Pr\left\{ \left|\check{\Theta}_{\rm p}  - \theta_{\mathrm{t}} \right|\ge\delta_2 \cup_{i=1}^u\left|\check{\varTheta}_i  - \vartheta_{\mathrm{t},i} \right| \ge\delta_2   \right\}\label{eq:probability_ub_unity}\\
 &\phantom{===}\le \lim_{n\to\infty} \left(\Pr\left\{ \left|\check{\Theta}_{\rm p}  - \theta_{\mathrm{t}} \right|\ge\delta_2\right\} + \sum_{i=1}^u\Pr\left\{ \left|\check{\varTheta}_i  - \vartheta_{\mathrm{t},i} \right| \ge\delta_2   \right\}\right) = 0 \label{eq:union bound} 
\end{align}
where \eqref{eq:probability_ub_unity} is because  probability $\Pr(\cdot)\leq 1$, inequality in \eqref{eq:union bound} is an application of the union bound, and equality in \eqref{eq:union bound} is due to conditions \ref{item: 1 pre} and \ref{item: 3 nuisance}.
The limit of the second term in \eqref{eq:consistency 2} is:
\begin{align}
   & \lim_{n\to\infty} \idotsint_{\Omega_{\delta_2}} p_{\check{\Theta}_{\rm p},\check{\vec{\varTheta}}}\left(\check{\theta}_{\rm p},\check{{\vartheta}}  \right) \Pr\left\{ \left|\check{\Theta}_{\rm r}\left( \check{\theta}_{\rm p},\check{\vec{\vartheta}} \right) - \theta_{\mathrm{t}} \right|>\epsilon_3 \right\} \dif \check{\theta}_{\rm p} \dif^u\check{\vec{\vartheta}}
 \nonumber \\
	&\phantom{===}\le    \lim_{n\to\infty} \Pr\left\{ \left|\check{\Theta}_{\rm p}  - \theta_{\mathrm{t}} \right|<\delta_2 \cap_{i=1}^u\left|\check{\varTheta}_i  - \vartheta_{\mathrm{t},i} \right| <\delta_2\right\} \max_{\{\check{\theta}_{\rm p},\check{\vec{\vartheta} }\}\in\Omega_{\delta_2}    } \Pr\left\{ \left|\check{\Theta}_{\rm r}\left( \check{\theta}_{\rm p},\check{\vec{\vartheta}} \right) - \theta_{\mathrm{t}} \right|>\epsilon_3 \right\} \nonumber \\
	&\phantom{===}\le   \lim_{n\to\infty}\max_{\{\check{\theta}_{\rm p},\check{\vec{\vartheta}}\}\in\Omega_{\delta_2}} \Pr\left\{ \left|\check{\Theta}_{\rm r}\left( \check{\theta}_{\rm p},\check{\vec{\vartheta}} \right) - \theta_{\mathrm{t}} \right|>\epsilon_3 \right\} = 0,\label{eq:consistency 2 up}
\end{align}
\end{widetext}
where the equality in \eqref{eq:consistency 2 up} is by condition \ref{condition:consistency}. 
Combining \eqref{eq:union bound} and \eqref{eq:consistency 2 up} results in $ \lim_{n\to\infty} \Pr\left\{ \left|\check{\Theta}_{\rm r} - \theta_{\mathrm{t}} \right|>\epsilon_3 \right\} = 0 $, showing the weak consistency of $ \check{\Theta}_{\rm r} $.

Next, we establish strong consistency using the almost-sure (a.s.) versions of conditions \ref{item: 1 pre}, \ref{item: 3 nuisance}, and \ref{condition:consistency}. Note that $\Check{\Theta}_{\rm r}$, $\Check{\Theta}_{\rm p}$, and  $\Check{\vec{\varTheta}}$ are functions of $n$. Let $ A = \left\{ \limsup_{n\to\infty} \left| \check{\Theta}_{\rm r}(\check{\Theta}_{\rm p},\check{\vec{\varTheta}})-\theta_{\mathrm{t}} \right|<\epsilon_3  \right\} $ and $ B = \left\{ \limsup_{n\to\infty} \left| \check{\Theta}_{\rm p}-\theta_{\mathrm{t}} \right|<\delta_2 \cap_{i=1}^u\left|\check{\varTheta}_i  - \vartheta_{\mathrm{t},i} \right| <\delta_2  \right\} $, where $\epsilon_3,\delta_2>0$. We need $\Pr\{A\}=1$ for strong consistency. By the law of total probability,
\begin{align}
    	\Pr\left\{A \right\} = \Pr\left\{A | B\right\}\Pr\left\{ B\right\}+\Pr\left\{A | B^c\right\}\Pr\left\{ B^c\right\},
\end{align}
where $B^c$ is the complement of $B$. Strong consistency follows as $\Pr\left\{A | B\right\}=1$ by condition \ref{condition:consistency} since $B$ is the event that $\check{\Theta}_{\rm p} $ and  $\check{\varTheta}_i $, for all $i$, are in the neighborhoods of $\theta_{\mathrm{t}}$ and the respective $\vartheta_{\mathrm{t},i}$ for infinitely many $n$, $\Pr\left\{ B\right\} =1$, $\Pr\left\{ B^c\right\} =0$ by conditions \ref{item: 1 pre} and \ref{item: 3 nuisance}, and $\Pr\left\{A | B^c\right\}\le 1$.

Finally, we prove the asymptotic normality of $\check{\Theta}_{\rm r}$ using weak consistency.
Define the following random variables: 
\begin{align}
Z_n\left(\check{\theta}_{\rm p},\check{\vec{\vartheta}}\right)&= \sqrt{n - f(n)} \left( \check{\Theta}_{\rm r}\left(\check{\theta}_{\rm p},\check{\vec{\vartheta}} \right) - \theta_{\mathrm{t}} \right)\\
Z_n&=  E_{\check{\Theta}_{\rm p},\check{\vec{\varTheta}}} \left[ Z_n\left(\check{\Theta}_{\rm p},\check{\vec{\varTheta}}\right)\right]\\
& = \sqrt{n - f(n)} \left( E_{\check{\Theta}_{\rm p},\check{\vec{\varTheta}}} \left[\check{\Theta}_{\rm r}\left(\check{\Theta}_{\rm p},\check{\vec{\varTheta}} \right)\right] - \theta_{\mathrm{t}} \right).
\end{align}
Since the random variable $\check{\Theta}_{\rm r}$ describing the outcomes of the refinement estimator is the expectation over the outcomes of preliminary and nuisance vector parameter estimators $E_{\check{\Theta}_{\rm p},\check{\vec{\varTheta}}} \left[\check{\Theta}_{\rm r}\left(\check{\Theta}_{\rm p},\check{\vec{\varTheta}} \right)\right]$, we need to show that 
\begin{align}
\lim_{n\to\infty}\left| F_{Z_n}\left(z\right) - \Phi\left( z \sqrt{\mathcal{J}_{\theta_{\mathrm{t}}}} \right) \right| = 0, \label{eq:normality theta0 proof}
\end{align}
where $ \Phi(x)=\frac{1}{\sqrt{2\pi}}\int_{-\infty}^x e^{-t^2/2}dt $ is the cumulative distribution function (c.d.f.) of $\mathcal{N}\left(0,1\right)$,
\begin{align}
F_{Z_n}\left(z\right) &= \!\!\!\idotsint\limits_{\Gamma\times\Gamma_1\times\dots\times\Gamma_u} p_{\Check{\Theta}_{\rm p},\check{\vec{\varTheta}}}\left(\Check{\theta}_{\rm p},\check{\vec{\vartheta}}\right)   F_{Z_n\left(\check{\theta}_{\rm p},\check{\vec{\vartheta}}\right)}\left(z\right) \dif \check{\theta}_{\rm p}\dif^u \check{\vec{\vartheta}}\nonumber\\
&= E_{\check{\Theta}_{\rm p},\check{\vec{\varTheta}}} \left[ F_{Z_n\left(\check{\Theta}_{\rm p},\check{\vec{\varTheta}}\right)}\left(z\right)    \right],
\end{align}
and $F_{Z_n\left(\check{\theta}_{\rm p},\check{\vec{\vartheta}}\right)}$ is the c.d.f.~of $Z_n\left(\check{\theta}_{\rm p},\check{\vec{\vartheta}}\right)$.
Using the triangle inequality, we have
\begin{widetext}
\begin{align}
         \left| F_{Z_n}\left(z\right) - \Phi\left( z\sqrt{\mathcal{J}_{\theta_{\mathrm{t}}} }\right) \right| &\le  \left| F_{Z_n}\left(z\right) -  E_{\check{\Theta}_{\rm p},\check{\vec{\varTheta}}}\left[ \Phi\left( z\sqrt{\mathcal{I}_{\theta_{\mathrm{t}},\check{\Theta}_{\rm p},\check{\vec{\varTheta}}}}\right)\right] \right| +\left| E_{\check{\Theta}_{\rm p},\check{\vec{\varTheta}}}\left[ \Phi\left( z\sqrt{\mathcal{I}_{\theta_{\mathrm{t}},\check{\Theta}_{\rm p},\check{\vec{\varTheta}}}}\right)\right] - \Phi\left( z\sqrt{\mathcal{J}_{\theta_{\mathrm{t}}} }\right) \right|\label{eq:hayashi conv in d upper}.
\end{align}
\end{widetext}
Consider the first term in \eqref{eq:hayashi conv in d upper},
\begin{align}
         &\left| F_{Z_n}\left(z\right) -  E_{\check{\Theta}_{\rm p},\check{\vec{\varTheta}}}\left[ \Phi\left( z\sqrt{\mathcal{I}_{\theta_{\mathrm{t}},\check{\Theta}_{\rm p},\check{\vec{\varTheta}}}}\right)\right] \right|  \nonumber\\
         &=  \left| E_{\check{\Theta}_{\rm p},\check{\vec{\varTheta}}}\left[ F_{Z_n\left( \check{\Theta}_{\rm p},\check{\vec{\varTheta}} \right)}\left(z\right) - \Phi\left( z \sqrt{\mathcal{I}_{\theta_{\mathrm{t}},\check{\Theta}_{\rm p},\check{\vec{\varTheta}} }} \right)\right] \right| \label{eq:up1 eq} \\
	& \le   E_{\check{\Theta}_{\rm p},\check{\vec{\varTheta}}}\left[ \left| F_{Z_n\left( \check{\Theta}_{\rm p},\check{\vec{\varTheta}} \right)}\left(z\right) - \Phi\left( z \sqrt{\mathcal{I}_{\theta_{\mathrm{t}},\check{\Theta}_{\rm p},\check{\vec{\varTheta}} }} \right)\right|\right]  \label{eq:up1 leq}\\
	&= \int_{\Omega_{\delta_2}^c} p_{\check{\Theta}_{\rm p},\check{\vec{\varTheta}}}\left( \check{\theta}_{\rm p},\check{\vec{\vartheta}} \right)  \nonumber\\
 &\phantom{\int}\times \left| F_{Z_n\left( \check{\Theta}_{\rm p},\check{\vec{\varTheta}} \right)}\left(z\right) - \Phi\left( z \sqrt{\mathcal{I}_{\theta_{\mathrm{t}},\check{\Theta}_{\rm p},\check{\vec{\varTheta}} }} \right)\right|    \dif^u \check{\vec{\vartheta}} \nonumber \\
 & +  \int_{\Omega_{\delta_2}} p_{\check{\Theta}_{\rm p},\check{\vec{\varTheta}}}\left( \check{\theta}_{\rm p},\check{\vec{\vartheta}} \right)   \nonumber\\
 &\phantom{\int}\times\left| F_{Z_n\left( \check{\Theta}_{\rm p},\check{\vec{\varTheta}} \right)}\left(z\right) \!- \! \Phi\left( z \sqrt{\mathcal{I}_{\theta_{\mathrm{t}},\check{\Theta}_{\rm p},\check{\vec{\varTheta}} }} \right)\right|   \dif \check{\theta}_{\rm p}\dif^u \check{\vec{\vartheta}}, \label{eq:up1 split }
\end{align}
where \eqref{eq:up1 eq} is by the definition of $ F_{Z_n}\left(z\right) $ and \eqref{eq:up1 leq} is from moving the absolute value inside the expectation.
We use the union bound and the fact that $ \left| F_{Z_n\left( \check{\Theta}_{\rm p},\check{\vec{\varTheta}} \right)}\left(z\right) - \Phi\left( z \sqrt{\mathcal{I}_{\theta_{\mathrm{t}},\check{\Theta}_{\rm p},\check{\vec{\varTheta}} }} \right)\right|\le 2 $ to upper bound the first term in \eqref{eq:up1 split } by $ 2 \left(\Pr\left\{| \check{\Theta}_{\rm p} - \theta_{\mathrm{t}} |>\delta_2\right\}+\sum_{i=1}^u\Pr\left\{\left|\check{\varTheta}_i  - \vartheta_{\mathrm{t},i} \right| >\delta_2 \right\}\right)$. Taking the limits as $n\to\infty$ yields zero by conditions \ref{item: 1 pre} and \ref{item: 3 nuisance}. The second term in \eqref{eq:up1 split } can be upper bounded by $ \left| F_{Z_n\left( \check{\theta}^{*}_{\rm p},\check{\vec{\vartheta}}^* \right)}\left(z\right) - \Phi\left( z \sqrt{\mathcal{I}_{\theta_{\mathrm{t}},\check{\theta}^{\ast}_{\rm p},\check{\vec{\vartheta}}^* }} \right)\right|$, where 
\begin{align} 
\{\theta^{*}_{\rm p},\check{\vec{\vartheta}}^*\} &= \arg \max_{\check{\theta}_{\rm p},\check{\vec{\vartheta}}\in\Omega_{\delta_2}} \left| F_{Z_n\left( \check{\theta}_{\rm p},\check{\vec{\vartheta}} \right)}\left(z\right) - \Phi\left( z \sqrt{\mathcal{I}_{\theta_{\mathrm{t}},\check{\theta}_{\rm p},\check{\vec{\vartheta}} }} \right)\right|.
\end{align}
By condition \ref{condition:asymptotic normality},
\begin{align}
\lim_{n\to\infty} \left| F_{Z_n\left( \check{\theta}^{*}_{\rm p},\check{\vec{\vartheta}}^* \right)}\left(z\right) - \Phi\left( z \sqrt{\mathcal{I}_{\theta_{\mathrm{t}},\check{\theta}^{\ast}_{\rm p},\check{\vec{\vartheta}}^* }} \right)\right| =0.
\end{align}
Thus, \eqref{eq:up1 split } yields
\begin{align}
       \lim_{n\to\infty}    \left| F_{Z_n}\left(z\right) -  E_{\check{\Theta}_{\rm p},\check{\vec{\varTheta}}}\left[ \Phi\left( z\sqrt{\mathcal{I}_{\theta_{\mathrm{t}},\check{\Theta}_{\rm p},\check{\vec{\varTheta}}}}\right)\right] \right| = 0. \label{eq:up1 interchanging}
\end{align}
For the second term in \eqref{eq:hayashi conv in d upper},
\begin{align}
   & \left| E_{\check{\Theta}_{\rm p},\check{\vec{\varTheta}}}\left[ \Phi\left( z\sqrt{\mathcal{I}_{\theta_{\mathrm{t}},\check{\Theta}_{\rm p},\check{\vec{\varTheta}}}}\right)\right] - \Phi\left( z\sqrt{\mathcal{J}_{\theta_{\mathrm{t}}} }\right) \right| \nonumber\\
&\le   E_{\check{\Theta}_{\rm p},\check{\vec{\varTheta}}}\left[\left| \Phi\left( z\sqrt{\mathcal{I}_{\theta_{\mathrm{t}},\check{\Theta}_{\rm p},\check{\vec{\varTheta}}}}\right) - \Phi\left( z\sqrt{\mathcal{J}_{\theta_{\mathrm{t}}} }\right) \right|\right],\label{eq:up2 interchanging}
\end{align}
where \eqref{eq:up2 interchanging} is from first moving $\Phi\left( z\sqrt{\mathcal{J}_{\theta_{\mathrm{t}}} }\right)$ and the the absolute value into the expectation.
Recall that $\mathcal{I}_{\theta_{\mathrm{t}},\theta_{\mathrm{t}},\vec{\vartheta_{\mathrm{t}}}}=\mathcal{J}_{\theta_{\mathrm{t}}} $.
Conditions \ref{item: 1 pre}, \ref{item: 3 nuisance}, and \ref{item:3 continuous} with continuous mapping theorem \cite[Th.~29.2]{billingsley95measure} imply that $\Phi\left( z\sqrt{\mathcal{I}_{\theta_{\mathrm{t}},\check{\Theta}_{\rm p},\check{\vec{\varTheta}} }}\right) \overset{p}{\to} \Phi\left( z\sqrt{\mathcal{J}_{\theta_{\mathrm{t}}} }\right) $. Since $  \left|\Phi\left( z\sqrt{\mathcal{I}_{\theta_{\mathrm{t}},\check{\Theta}_{\rm p},\check{\vec{\varTheta}} }}\right) \right| \le1$, $\Phi\left( z\sqrt{\mathcal{I}_{\theta_{\mathrm{t}},\check{\Theta}_{\rm p},\check{\vec{\varTheta}} }}\right) $ is uniformly integrable. By Vitali convergence theorem \cite[Corr.~to Th.~16.14]{billingsley95measure}, the limit of \eqref{eq:up2 interchanging} yields:
\begin{align}
    \lim_{n\to\infty}E_{\check{\Theta}_{\rm p}}\left[ \left|\Phi\left( z\sqrt{\mathcal{I}_{\theta_{\mathrm{t}},\check{\Theta}_{\rm p},\check{\vec{\varTheta}} }}\right) - \Phi\left( z\sqrt{\mathcal{J}_{\theta_{\mathrm{t}}} }\right) \right|\right]=0. \label{eq:tri 2}
\end{align}
Combining \eqref{eq:hayashi conv in d upper}, \eqref{eq:up1 interchanging}, and \eqref{eq:tri 2} yields $  \lim_{n\to\infty}\left|  F_{Z_n}\left(x\right) - \Phi\left( z\sqrt{\mathcal{J}_{\theta_{\mathrm{t}}} }\right) \right| =0 $ and asymptotic normality in \eqref{eq:asymptotic normality}.
\end{proof}

\section{Strong Consistency of the Preliminary Transmittance and Phase Shift Estimators} \label{app:pre consistency}

The output of the heterodyne measurement is a sequence of pairs $ \{a_i,b_i \}_{i=1}^{f(n)} $ that is described by i.i.d.~Gaussian random variables $ A_i\sim\mathcal{N}\left(\sqrt{\theta_{\mathrm{t}} \bar{n}_{\rm S}}\cos\gamma_{\mathrm{t}}, \bar{n}_{\rm B}+\frac{1}{2} \right)  $ and $ B_i\sim\mathcal{N}\left(\sqrt{\theta_{\mathrm{t}} \bar{n}_{\rm S}}\sin\gamma_{\mathrm{t}}, \bar{n}_{\rm B}+\frac{1}{2} \right)$, where $ \gamma_{\mathrm{t}} $ and $ \theta_{\mathrm{t}} $ are the true values of $ \gamma $ and $ \theta $ respectively, $ f(n) $ satisfies $ \lim\limits_{n\to\infty}f(n) \to \infty $ and $ \lim\limits_{n\to \infty} \frac{f(n)}{n} = 0 $, and the joint p.d.f.~of $ \{A_i,B_i \} $ is \cite{weedbrook12gaussianQIrmp},\cite[Ch.~7.3.2]{guha04mastersthesis}:
	\begin{align}
	p(a_i,b_i;\theta_{\mathrm{t}},\gamma_{\mathrm{t}}) = \frac{e^{-\frac{(a_i-\sqrt{\theta_{\mathrm{t}} \bar{n}_{\rm S} } \cos\gamma_{\mathrm{t}})^2+(b_i-\sqrt{\theta_{\mathrm{t}} \bar{n}_{\rm S} } \sin\gamma_{\mathrm{t}})^2}{2\bar{n}_{\rm B}+1}}}{\pi(2\bar{n}_{\rm B}+1) }.\label{eq:pre p}
	\end{align} 
	The maximum likelihood estimators $ \check{\theta}\left(f(n)\right) $ and $ \check{\gamma}\left(f(n)\right) $ satisfy:
	\begin{align}
	\left(\partial_{\theta}\sum_{i=1}^{f(n)} \log p(a_i,b_i;\theta,\gamma)  \right) \Big|_{\theta = \check{\theta}\left(f(n)\right), \gamma = \check{\gamma}\left(f(n)\right)  }  = 0\label{eq:MLE 1}\\
	\left(\partial_{\gamma}\sum_{i=1}^{f(n)} \log p(a_i,b_i;\theta,\gamma)  \right) \Big|_{\theta = \check{\theta}\left(f(n)\right), \gamma = \check{\gamma}\left(f(n)\right)  }  = 0.\label{eq:MLE 2}
	\end{align}
	Substituting $ \eqref{eq:pre p} $ into $ \eqref{eq:MLE 1} $ and $ \eqref{eq:MLE 2} $ yields:
	\begin{align}
	\frac{1}{f(n)} \sum_{i=1}^{f(n)} a_i \cos\check{\gamma}\left(f(n)\right) + b_i\sin\check{\gamma}\left(f(n)\right)& = \sqrt{\check{\theta}\left(f(n)\right)\bar{n}_{\rm S} }\\
	\frac{1}{f(n)} \sum_{i=1}^{f(n)} a_i \cos\check{\gamma}\left(f(n)\right) - b_i\sin\check{\gamma}\left(f(n)\right)& =0.
	\end{align}
	Therefore,
	\begin{align}
	\check{\theta}\left(f(n)\right)& = \left( \bar{A}\left(f(n)\right)\right)^{2}+\left (\bar{B}\left(f(n)\right)\right)^{2} \\
	\check{\gamma}\left(f(n)\right)& = \arctan \frac{\bar{B}\left(f(n)\right)}{\bar{A}\left(f(n)\right)},
	\end{align}
	where $ \bar{A}\left(f(n)\right)=\frac{1}{f(n)\sqrt{\bar{n}_{\rm S}}} \sum_{i=1}^{f(n)} A_i$, and $ \bar{B}\left(f(n)\right)=\frac{1}{f(n)\sqrt{\bar{n}_{\rm S}}} \sum_{i=1}^{f(n)} B_i $. 
	By Khinchine's strong law of large numbers (SLLN) \cite[Th.~10.13]{folland1999real}, we have:
	\begin{align}
	\bar{A}\left(f(n)\right)&\xrightarrow{a.s.}\sqrt{\theta_{\mathrm{t}}} \cos\gamma_{\mathrm{t}} \\
	\bar{B}\left(f(n)\right)&\xrightarrow{a.s.} \sqrt{\theta_{\mathrm{t}}} \sin\gamma_{\mathrm{t}}.
	\end{align}
	Since $g(x)=x^2$ is a continuous function of $x$, by the continuous mapping theorem \cite[Th.~25.7]{billingsley95measure} we have:
	\begin{align}
	\left( \bar{A}\left(f(n)\right)\right)^2&\xrightarrow{a.s.}\theta_{\mathrm{t}} \cos^2\gamma_{\mathrm{t}} \\
	\left ( \bar{B}\left(f(n)\right)\right)^2&\xrightarrow{a.s.}\theta_{\mathrm{t}} \sin^2\gamma_{\mathrm{t}}.
	\end{align}
	Therefore,
	\begin{align}
	\check{\theta}\left(f(n)\right) = \bar{A}\left(f(n)\right)^{2}+\bar{B}\left(f(n)\right)^{2} \xrightarrow{a.s.}  \theta_{\mathrm{t}},
	\end{align}
	and, since $ \arctan(x) $ is continuous, by continuous mapping theorem,
	\begin{align}
	\check{\gamma}\left(f(n)\right) &= \arctan \frac{\bar{B}\left(f(n)\right)}{\bar{A}\left(f(n)\right)} \xrightarrow{a.s.}  \arctan \frac{\sqrt{\theta_{\mathrm{t}}} \sin\gamma_{\mathrm{t}}}{\sqrt{\theta_{\mathrm{t}}} \cos\gamma_{\mathrm{t}}} = \gamma_{\mathrm{t}}.
	\end{align}

\section{Derivation of \eqref{eq:likelihood function} and \eqref{eq:summand}}
\label{app:pmf}
Applying the channel's phase shift $\hat{U}_R(\gamma_{\rm t})$ to the Fock-diagonal representation of the output state without the phase shift derived in \cite[App.~B.A]{gong22losssensing} yields the output state:
\begin{align}
    \hat{\sigma}_{IR} = \sum_{s,t=0}^\infty r_{s,t}\hat{U}_{R}(\gamma_{\rm t})\hat{S}^{\dagger}_{IR}(\zeta)\ket{s,t}\bra{s,t}\hat{S}_{IR}(\zeta)\hat{U}_{R}^{\dagger}(\gamma_{\rm t}),\label{eq:output state}
\end{align}
where $r_{s,t}$ is in \eqref{eq:rst}, and $\zeta$ is a function of $\theta_{\rm t}$, $\bar{n}_{\rm S}$, and $\bar{n}_{\rm B}$ given in \cite[Eqs.~(56)]{gong22losssensing}.
After applying the phase shift compensator $\hat{U}_{R}(\check{\gamma})$ and the two-mode squeezer $\hat{S}_{IR}(\omega)$ in the receiver, the p.m.f.~for the output of PNR detectors is: 
\begin{align}
    & p_{K,M}(k,m;\theta,\check{\theta}_{\rm p}, \check{\gamma}) \nonumber \\
    &=\bra{k,m}\hat{S}^{\dagger}_{IR}(\omega)\hat{U}_R\left(-\check{\gamma}\right)\hat{\sigma}_{IR}\hat{U}_R\left(\check{\gamma}\right)\hat{S}_{IR}(\omega) \ket{k,m}   \\
 &=   \sum_{s,t=0}^{\infty} r_{s,t}|\bra{s,t}\hat{S}_{IR}(\zeta)\hat{U}_R(\check{\gamma}-\gamma_{\rm t})\hat{S}_{IR}(\omega)\ket{k,m} |^2.\label{eq:pmf 1}
\end{align}
First, consider $ \hat{S}_{IR}(\omega)$ acting on $\ket{k,m}$. By \cite[App.~C]{gong22losssensing}:
\begin{widetext}
\begin{align}
\hat{S}_{IR}(\omega)\ket{k,m}&= \sum_{i_1=0}^{\min(k,m)} \sum_{a_1=0}^{\infty} \frac{(-\tau_1)^{a_1}(\tau_1^*)^{i_1}\sqrt{k!m!(k-i_1+a_1)!(m-i_1+a_1)!}}{a_1!i_1!(k-i_1)!(m-i_1)!\nu_1^{k + m - 2i_1 + 1}}\ket{k-i_1+a_1,m-i_1+a_1}, \label{eq:pmf o1} 
\end{align}
\end{widetext}
where $ \tau_1 = \tanh\omega$, and $\nu_1 = \cosh\omega$.
Next, consider $\hat{U}_R(\check{\gamma}-\gamma_{\rm t})$ acting on $\ket{k-i_1+a_1,m-i_1+a_1}$ in \eqref{eq:pmf o1}. By \cite[Eq.~(5.46)]{agarwal12quantumoptics}:
\begin{align}
&\hat{U}_R(\check{\gamma}-\gamma_{\rm t})\ket{k-i_1+a_1,m-i_1+a_1} \nonumber \\
      &=e^{-j(\check{\gamma}-\gamma_{\rm t})(k-i_1+a_1)} \ket{k-i_1+a_1,m-i_1+a_1}\label{eq:pmf o2}
\end{align}
Then, consider $\hat{S}_{IR}(\zeta)$ acting on $ \ket{k-i_1+a_1,m-i_1+a_1}$:
\begin{widetext}
\begin{align}
&\hat{S}_{IR}(\zeta)\ket{k-i_1+a_1,m-i_1+a_1}\nonumber \\ 
&\phantom{===}=\sum_{i_2=0}^{i_{\rm u}} \sum_{a_2=0}^{\infty} \frac{(-\tau_2)^{a_2} (\tau_2^*)^{i_2}\sqrt{(k-i_1+a_1)!(m-i_1+a_1)!(k-i_1+a_1-i_2+a_2)!(m-i_1+a_1-i_2+a_2)!}}{a_2!i_2!(k-i_1+a_1-i_2)!(m-i_1+a_1-i_2)!\nu_2^{k + m -2i_1 +2 a_1 - 2i_2 + 1}}\nonumber\\
&\phantom{====\sum_{i_2=0}^{i_{\rm u}} \sum_{a_2=0}^{\infty}}\times\ket{k-i_1+a_1-i_2+a_2,m-i_1+a_1-i_2+a_2},\label{eq:pmf o3}
\end{align}
\end{widetext}
where $\tau_2 = \tanh\zeta$,  $\nu_2 = \cosh \zeta$, and $i_{\rm u}=\min(k-i_1+a_1,m-i_1+a_1)$.
Finally, evaluating $\left\langle s,t \middle| k-i_1+a_1-i_2+a_2,m-i_1+a_1-i_2+a_2\right\rangle$ yields:
\begin{align}
    a_2 &= s-k+i_1-a_1+i_2, \label{eq:pmf o4}\\
    t &= s-k+m. \label{eq:pmf o5}
\end{align}
By \eqref{eq:pmf o4} and $a_2>0 $, the lower limit of $i_2$ is $ i_{\rm l}=\max(0,k-s-i_1+a_1)$. Combining \eqref{eq:pmf 1}--\eqref{eq:pmf o5} yields \eqref{eq:likelihood function} and \eqref{eq:summand}.

\section{Lemmas for Analysis of MLE Applied to Optimal Transmittance Measurement Outcomes}
\label{app:lemmas_mle}
Here, we present the results we use to demonstrate the strong consistency and asymptotic normality of the quantum-optimal transmittance sensor described in Section \ref{sec:transmittance sensing}.
We rely on the strong uniform law of large numbers to obtain strong consistency.
To state it, we need to introduce separability:
\begin{definition}[Separability \protect{\cite[Assumption~A-1]{huber1967behavior}}]
\label{def:separability}
For every $\theta \in \Gamma$, a measurable function $f(x,\theta)$ is \emph{separable} if there is a null set $\mathcal{U}$ and a countable subset $\Gamma^{\prime}\subset\Gamma$ such that, for every open set $ \mathcal{T} \subset \Gamma$ and every closed interval $B$, the sets $\left\{x|f(x,\theta)\in B, \forall \theta \in \mathcal{T}\right\}$ and $\left\{x|f(x,\theta)\in B, \forall \theta \in \mathcal{T}\cap \Gamma^{\prime} \right\}$ differ by at most a subset of $\mathcal{U}$.
\end{definition}
\begin{remark}[\hspace{1sp}\protect{\cite[Ex.~38.3]{billingsley95measure}}]\label{remark:separability}
$f(x;\theta)$ is separable if it is continuous in $x$ for all $\theta$.
\end{remark}
\begin{lemma}[Strong uniform law of large numbers \protect{\cite[Lem.~1]{tauchen1985diagnostic}}]\label{lemma:sulln}
Let  $\left\{f(X_i,\theta)\right\}_{i=1}^{n}$ be a function of a sequence of $n$ i.i.d.~random variables $\left\{X_i\right\}_{i=1}^{n}$. Then $ \frac{1}{n}\sum_{i=1}^{n}f(X_i,\theta) \xrightarrow{a.s.} E[f(X_1,\theta)]$ uniformly in $\theta$, if:
\begin{enumerate}
\item \label{item: measurable} $ f(x,\theta) $ is measurable in $ x $ for each $ \theta \in \Gamma $.
\item  \label{item: separable} $f(x,\theta)$ is separable per Definition \ref{def:separability};
\item \label{item:continuity} $f(x,\theta)$ is continuous almost everywhere: for all $\theta$ $\Pr\left(\left\{ x:\lim_{\gamma\to\theta}f(x,\gamma) = f(x,\theta) \right\}\right)=1$.
\item\label{item:dominate}  $ f(X,\theta) $ is dominated by $ g(X) $ for all $ \theta $ such that $ E[g(X)]<\infty $.
\end{enumerate}
\end{lemma}

Consider observations $\{x_i\}_{i=1}^{n}$ described by a sequence of $n$ i.i.d.~random variables $\left\{X_i\right\}_{i=1}^{n}$ with mass function $p_X(x;\theta)$.
Interpreting $p_X(x;\theta)$ as a likelihood function, the following lemma provides the conditions for the strong consistency of an MLE $\check{\theta}\left(\{x_i\}_{i=1}^{n}\right)\equiv\max_{\theta\in\Gamma}\frac{1}{n}\sum_{i=1}^{n}\log p_X(x_i;\theta)$:
\begin{lemma}[Strong consistency of the MLE \protect{\cite[Th.~1]{tauchen1985diagnostic}}]\label{lemma:strong_consistency}
$\check{\theta}\left(\left\{X_i\right\}_{i=1}^{n}\right)\xrightarrow{a.s.}\theta_{\rm t}$ if
\begin{enumerate}
\item \label{item:unique_maximum} $E[\log p_X(X;\theta)]$ has a unique maximum in $\theta$, and
\item \label{item:sulln} $\log p(x;\theta)$ satisfies the conditions for strong uniform law of large numbers in Lemma \ref{lemma:sulln}.
\end{enumerate}
\end{lemma}

We also need the following adaptation of \cite[Th.~4]{talvila2001necessary} (stated as \cite[Th.~3]{stevecheng}), to show regularity conditions in our proof of asymptotic normality:
\begin{lemma}\label{lemma: interchange} Let $\mathcal{X}$ be an open subset of $ \mathbb{R} $, and $\mathcal{W}$ be a measure space. Suppose $ f:\mathcal{X}\times \mathcal{W} \to \mathbb{R} $ satisfies the following conditions
\begin{enumerate}
\item \label{item:derivative lemma measurability condition} $f(x,w)$ is a measurable function of $ x $ and $ w $ jointly, and is integrable over $ w $, for almost all $ x\in \mathcal{X} $.
\item \label{item:derivative lemma absolute continuity condition}  For almost all $ w\in \mathcal{W} $, $ f(x,w) $ is an absolutely continuous function (see definition in \cite[Eq.~(3.31)]{folland1999real}) of $ x $.
\item \label{item:derivative lemma integral condition} For all compact intervals $ [a,b] \subseteq \mathcal{X} $:
\begin{align}
\int_{a}^{b}\int_{\mathcal{W}} \left| \partial_{x}f(x,w)  \right| \dif w\dif x <\infty.
\end{align}
\end{enumerate}
Then, $ \int_{\mathcal{W}} f(x,w) \dif w $ is an absolutely continuous function of $ x $, and, for almost all $ x\in \mathcal{X} $, its derivative exists and is:
\begin{align}
\partial_{x}\int_{\mathcal{W}} f(x,w)\dif w = \int_{\mathcal{W}}\partial_{x} f(x,w)\dif w.\label{eq:inter}
\end{align}
\end{lemma}
For completeness, we adapt the proof of \cite[Th.~4]{talvila2001necessary}:
\begin{proof}[Proof (Lemma \ref{lemma: interchange})] For almost all $ x\in \mathcal{X} $ and $ w \in \mathcal{W} $,
\begin{align}
&\partial_{x}\int_{\mathcal{W}}f(x,w)\dif w \nonumber\\
\qquad &=\lim\limits_{h\to0} \frac{1}{h}\left[\int_{\mathcal{W}}f(x+h,w)-f(x,w)\dif w\right] \label{eq:interchange proof 1}  \\
&=\lim\limits_{h\to0} \frac{1}{h}\left[\int_{\mathcal{W}} \int_{x}^{x+h}\partial_{x} f(x,w) \dif x\dif w\right] \label{eq:interchange proof 2} \\
&=\lim\limits_{h\to0} \frac{1}{h}\left[\int_{x}^{x+h}\int_{\mathcal{W}} \partial_{x} f(x,w) \dif w\dif x\right] \label{eq:interchange proof 3}\\
&=\frac{\dif}{\dif x} G(x),\label{eq:interchange proof 4}
\end{align}
where \eqref{eq:interchange proof 1} is by condition \ref{item:derivative lemma measurability condition} and the definition of derivative, \eqref{eq:interchange proof 2} is by condition \ref{item:derivative lemma absolute continuity condition} and the fundamental theorem of calculus for Lebesgue integral (FTCL) \cite[Th.~3.35]{folland1999real}, \cite[Th.~2]{talvila2001necessary}, and \eqref{eq:interchange proof 3} is by condition \ref{item:derivative lemma integral condition} and Fubini-Tonelli theorem \cite[Sec.~18]{billingsley95measure}.
Let $g(x)\equiv\int_{\mathcal{W}} \partial_{x} f(x,w) \dif w$ and $G(x)\equiv\int_a^x g(y)\dif y$ for some arbitrary $a\in\mathcal{X}$. Then \eqref{eq:interchange proof 4} follows by the definition of derivative and rearrangement of terms. The lemma follows from condition \ref{item:derivative lemma integral condition} and FTCL.
\end{proof}

\section{Strong Consistency of MLE Applied to Optimal Transmittance Measurement Outcomes}
\label{app:consistency}
The expectations here are with respect to $(K,M)$; thus, we drop the subscript $KM$ from $ E_{KM}[\cdot]$ for brevity.
We employ Lemma \ref{lemma:strong_consistency} to show that the MLE on the outcomes of our optimal transmittance measurement is strongly consistent. First, condition \ref{item:unique_maximum} is met: $E[\log p(K,M;\theta,\check{\theta}_{\rm p},\check{\gamma})]$ has a unique maximum per \cite[Lemma~2.2]{newey94largesampleestimation} as $p(k,m;\theta_1,\check{\theta}_{\rm p},\check{\gamma})\ne p(k,m;\theta_2,\check{\theta}_{\rm p},\check{\gamma})$ when $\theta_1\ne\theta_2$, and, per Section \ref{sec:transmittance sensing}, parameter space $\Gamma\equiv[\theta^{\prime\prime},1]$, with $\theta^{\prime\prime}>0$, is compact.

Meeting condition \ref{item:sulln} of Lemma \ref{lemma:strong_consistency} requires ensuring that the conditions of Lemma \ref{lemma:sulln} hold for $\log p(k,m;\theta,\check{\theta}_{\rm p},\check{\gamma})$.
Conditions \ref{item: measurable}–\ref{item:continuity} of Lemma \ref{lemma:sulln} are satisfied by inspection: $\log p(k,m;\theta,\check{\theta}_{\rm p},\check{\gamma})$ is measurable since it is a continuous function on a discrete measure, and $ \log p(k,m;\theta,\check{\theta}_{\rm p},\check{\gamma})$ is separable per Remark \ref{remark:separability}, since it is continuous in $(k,m)$ for each $ \theta \in \Gamma $ with factorials generalized to Gamma functions.

To verify condition \ref{item:dominate} of Lemma \ref{lemma:sulln}, we define the dominating function $\sup_{\theta\in\Gamma}\left|\log(p(K,M;\theta,\check{\theta}_{\rm p},\check{\gamma}))\right|$, and show that:
\begin{align}
E\left[\sup_{\theta\in\Gamma}\left|\log(p(K,M;\theta,\check{\theta}_{\rm p},\check{\gamma}))\right|\right]<\infty.\label{eq:MLE condition}
\end{align}
We upper-bound $\left|\log\left(p(k,m;\theta,\check{\theta}_{\rm p},\check{\gamma}\right)\right|$ using a lower bound on $p\left(k,m;\theta,\check{\theta}_{\rm p},\check{\gamma}\right)$, as it is in $(0,1]$.  Since each term in the summation over $s$ defining $p\left(k,m;\theta,\check{\theta}_{\rm p},\check{\gamma}\right)$ in \eqref{eq:likelihood function} is non-negative, any of them yields a lower bound.
When $ k\le m $, let $ s = 0 $. Then we have $ i_2 = k-i_1+a_1 $, and $t^{\prime} = m-k$, and
\begin{widetext}
\begin{align}
& \bra{0,m-k}\hat{S}_{IR}(\omega)\hat{U}_R(\check{\gamma}-\gamma_{\rm t})\hat{S}_{IR}(\zeta)\ket{k,m}  \nonumber \\
&\phantom{===}= \sqrt{\frac{k!m!}{(m-k)!}}\sum_{i_1=0}^{k} \sum_{a_1=0}^{\infty} \frac{(-\tau_1)^{a_1}\tau_1^{i_1}\nu_1^{-(k + m - 2i_1 + 1)}(m-i_1+a_1)!e^{-j(\check{\gamma}-\gamma_{\rm t})(k-i_1+a_1)}  \tau_2^{k-i_1+a_1}\nu_2^{-(m - k + 1)}}{a_1!i_1!(k-i_1)!(m-i_1)!}  \nonumber \\
&\phantom{===} = \sqrt{\binom{m}{k}} \frac{\left(\frac{\nu_2\tau_2}{\nu_1}+\nu_1\nu_2\tau_1(e^{j(\check{\gamma}-\gamma_{\rm t})}+\tau_1\tau_2)\right)^k}{e^{j(k-1)(\check{\gamma}-\gamma_{\rm t})}\left(\nu_1\nu_2(1+e^{-j(\check{\gamma}-\gamma_{\rm t})}\tau_1\tau_2)\right)^{m+1}}.\label{eq:eval_bound}
\end{align}
\end{widetext}
Substitution of \eqref{eq:eval_bound} into \eqref{eq:likelihood function} yields:
\begin{widetext}
\begin{align}
p(k,m;\theta,\check{\theta}_{\rm p},\check{\gamma})&\ge \frac{\binom{m}{k} r{0,m-k}\left|\frac{\nu_2\tau_2}{\nu_1}+\nu_1\nu_2\tau_1(e^{j(\check{\gamma}-\gamma_{\rm t})}+\tau_1\tau_2)\right|^{2k}}{(\nu_1\nu_2)^{2m+2}|1+e^{-j(\check{\gamma}-\gamma_{\rm t})}\tau_1\tau_2|^{2m} |e^{j(\check{\gamma}-\gamma_{\rm t})}+\tau_1\tau_2|^2} \geq r_{0,m-k} l_1^k l_2^{-m-1} = r_{0,m-k} l_3^k l_2^{-m+k-1}, \label{eq:pkm ks}
\end{align}
\end{widetext}
where the second inequality in \eqref{eq:pkm ks} is because the binomial coefficient $\binom{m}{k}=\frac{m!}{k!(m-k)!}\geq 1$, and
\begin{widetext}
\begin{align}
l_1 &\equiv\left|\frac{\nu_2\tau_2}{\nu_1}+\nu_1\nu_2\tau_1(e^{j(\check{\gamma}-\gamma_{\rm t})}+\tau_1\tau_2)\right|^{2}=\left(\frac{\nu_2\tau_2}{\nu_1}+\nu_1\nu_2\tau_1^2\tau_2\right)^2+ \nu_1^2\nu_2^2\tau_1^2+2\tau_1\nu_2^2\tau_2(1 +\nu_1^2\tau_1^2) \cos(\check{\gamma}-\gamma_{\rm t}).\\
l_2 &\equiv\nu_1^2\nu_2^2|1+e^{-j(\check{\gamma}-\gamma_{\rm t})}\tau_1\tau_2|^{2}= \nu_1^2\nu_2^2|e^{j(\check{\gamma}-\gamma_{\rm t})}+\tau_1\tau_2|^2 =\nu_1^2\nu_2^2(1+\tau_1^2\tau_2^2+2\tau_1\tau_2\cos(\check{\gamma}-\gamma_{\rm t}))\\
l_3 &\equiv\frac{l_1}{l_2}= \frac{\tau_1^2+2\tau_1\tau_2\cos(\check{\gamma}-\gamma_t)+\tau_2^2}{1+\tau_1^2\tau_2^2+2\tau_1\tau_2\cos(\check{\gamma}-\gamma_t)}.
\end{align}
\end{widetext}
Now, note that
\begin{align}
l_2 &\ge\nu_1^2\nu_2^2(1+\tau_1^2\tau_2^2-2|\tau_1| |\tau_2|)\nonumber\\
&= \nu_1^2\nu_2^2 (1-|\tau_2||\tau_1|)^2  \nonumber\\
&= \left( \cosh(|\omega|)\cosh(|\zeta|)-\sinh(|\omega|)\sinh(|\zeta|) \right)^2\nonumber\\
&= \cosh^2\left(|\omega| - |\zeta|\right) \ge 1.
\end{align}
Furthermore, comparison of the denominator and the numerator of $l_3$ yields $l_3\in\left(0,1\right)$, as:
\begin{multline}
\tau_1^2+2\tau_1\tau_2\cos(\check{\gamma}-\gamma_t)+\tau_2^2\\
-\left(1+\tau_1^2\tau_2^2+2\tau_1\tau_2\cos(\check{\gamma}-\gamma_t)\right)=-\frac{1}{\nu_1^2\nu_2^2}<0.
\end{multline}
When $ k\ge m $, let $ s = k-m $.
We have
$i_2=m-i_1+a_1$, $ t^{\prime} = 0$, and calculations similar to those in \eqref{eq:eval_bound} and \eqref{eq:pkm ks} yield:
\begin{align}
p(k,m;\theta,\check{\theta}_{\rm p},\check{\gamma})&\ge r{k-m,0}l_2^{-k-1} l_1^{m}\nonumber\\
&= r_{k-m,0} l_3^m l_2^{-k+m-1} \label{eq:pkm ms}.
\end{align}
Combining \eqref{eq:pkm ks} and \eqref{eq:pkm ms}, for all $ k $ and $ m $, we have:
\begin{align}
p(k,m;\theta,\check{\theta}_{\rm p},\check{\gamma})\ge r{k,m} l_3^{k+m} l_2^{-k-m-1}.
\end{align}
As $ 0<p(k,m;\theta,\check{\theta}_{\rm p},\check{\gamma})<1 $,
\begin{align}
&E\left[\sup_{\theta\in\Gamma}\left|\log(p(K,M;\theta,\check{\theta}_{\rm p},\check{\gamma}))\right|\right] \nonumber \\
&\leq  - E\left [ \log r{K,M}\right  ] -E\left [ -(K+M+1)\log l_2^{\ast} \right ]\nonumber\\
& \phantom{- E\left [ \log r_{K,M}\right  ]-(K+M+1)}-E\left [ (K+M)\log l_3^{\ast} \right ] \nonumber \\
&\leq E\left [ K\right  ] L_{1}^\ast  + E\left [ M\right  ] L_{2}^\ast  + E\left [(K+M+1) \right ]\log l_2^{\ast}\nonumber\\
& -E\left [ K+M \right ]\log l_3^{\ast}+\sup_{\theta\in\Gamma}\log\left((1+N_1)(1+N_2)\right),\label{eq:expectation of loglikelihood}
\end{align}
where $r_{k,m}$ is defined in \eqref{eq:rst} and
\begin{align}
l_2^{\ast}&=\sup_{\theta\in\Gamma}l_2, & l_3^{\ast}&=\inf_{\theta\in\Gamma}l_3\nonumber\\
L_{1}^\ast&=\sup_{\theta\in\Gamma}\log(1+1/N_1), & L_{2}^\ast&=\sup_{\theta\in\Gamma}\log(1+1/N_2).\nonumber
\end{align}

Since the energy of the received photons is finite, $ E[K] $ and $ E[M] $ are finite for all $ \theta$. Thus, every term in \eqref{eq:expectation of loglikelihood} is finite, satisfying \eqref{eq:MLE condition} and resulting in strong consistency: $\check{\Theta}_{\rm r}\left( \check{\theta}_{\rm p},\check{\gamma} \right) \xrightarrow{a.s.}\theta_{\mathrm{t}}$.

\section{Asymptotic Normality of MLE Applied to Optimal Transmittance Measurement Outcomes}\label{app:normality proof}

Again, the expectations here are with respect to $(K,M)$, and we drop the subscript $KM$ from $ E_{KM}[\cdot]$.
First, for $\theta \in \Gamma\equiv[\theta^{\prime\prime},1] $, we show that the following regularity conditions hold in Appendix  \ref{app: normality}:
\begin{align}
    \sum_{km} \partial_{\theta} p(k,m;\theta,\check{\theta}_{\rm p},\check{\gamma}) &= \partial_{\theta}\sum_{km}  p(k,m;\theta,\check{\theta}_{\rm p},\check{\gamma}) =0  \label{eq:interchange 1} \\
    \sum_{km} \partial^2_{\theta} p(k,m;\theta,\check{\theta}_{\rm p},\check{\gamma}) &= \partial^2_{\theta}\sum_{km} p(k,m;\theta,\check{\theta}_{\rm p},\check{\gamma})=0  \label{eq:interchange 2}. 
\end{align}
Furthermore, $\partial_{\theta} p(k,m;\theta,\check{\theta}_{\rm p},\check{\gamma}) $ and $\partial^2_{\theta} p(k,m;\theta,\check{\theta}_{\rm p},\check{\gamma}) $ exist and are continuous for all $\theta, \check{\theta}_{\rm p}\in \left[\theta^{\prime\prime},1\right]$ and $\check{\gamma}\in[-\pi,\pi]$. By \cite[Prop.~IV.C.4]{poor94intro-det-est}, \eqref{eq:interchange 1} and \eqref{eq:interchange 2} yield:
\begin{align}
    \mathcal{I}_{\theta_{\mathrm{t}},\check{\theta}_{p},\check{\gamma}}&= E\left[\left(  \partial_{\theta} \log  p(K,M;\theta,\check{\theta}_{p},\check{\gamma}) \right)^2\Big|_{\theta=\theta_{\mathrm{t}}}\right] \\
    &=- E\left[ \left(\partial^2_{\theta} \log  p(K,M;\theta,\check{\theta}_{p},\check{\gamma}) \right)\Big|_{\theta=\theta_{\mathrm{t}}}\right]
\end{align}
By the definition of the MLE, we select $ \check{\Theta}_{\rm r}\left(\check{\theta}_{\rm p},\check{\gamma}\right) $ such that
\begin{align}
    \frac{1}{n^\prime}\sum_{i}\partial_{\theta}\log p(K_i,M_i;\theta,\check{\theta}_{\rm p},\check{\gamma})\Big|_{\theta = \check{\Theta}_{\rm r}\left(\check{\theta}_{\rm p},\check{\gamma}\right)} = 0, \label{eq:d of MLe}
\end{align}  
where the summations over $i$ in this appendix are from $i=f(n)+1$ to $i=n$, and $n^{\prime}\equiv n-f(n)$.
Application of the mean value theorem to \eqref{eq:d of MLe} yields:
\begin{multline}
    \sqrt{n^\prime}\left( \check{\Theta}_{\rm r}\left(\check{\theta}_{\rm p},\check{\gamma}\right) - \theta_{\mathrm{t}} \right) \\
    = \frac{\frac{1}{\sqrt{n^\prime}} \sum_{i}\left(\partial_{\theta}\log p\left(K_i,M_i;\theta,\check{\theta}_{\rm p},\check{\gamma}\right)\right)\Big|_{\theta=\theta_{\mathrm{t}}}}{-\frac{1}{n^\prime}\sum_{i} \left(\partial^2_{\theta}\log p\left(K_i,M_i; \theta,\check{\theta}_{\rm p},\check{\gamma}\right)\right)\Big|_{\theta=\bar{\Theta}}}, \label{eq: ratio}
\end{multline}
where $\left|\bar{\Theta} -\theta_{\mathrm{t}} \right|\le \left|\check{\Theta}_{\rm r}\left(\check{\theta}_{\rm p},\check{\gamma}\right) -\theta_{\mathrm{t}} \right| $. Consistency of $\check{\Theta}_{\rm r}\left(\check{\theta}_{\rm p},\check{\gamma}\right)$ shown in Appendix \ref{app:consistency} yields:
\begin{align}
    \bar{\Theta} \overset{p}{\to} \theta_{\mathrm{t}}.\label{eq: bar converge}
\end{align}
Now we adapt the standard results (e.g., from \cite[Prop.~IV.D.2]{poor94intro-det-est}).
Consider the numerator in \eqref{eq: ratio}. 
By \eqref{eq:interchange 1}, we have $E\left[  \left(\partial_{\theta}\log p\left(K,M;\theta,\check{\theta}_{\rm p},\check{\gamma}\right)\right)\Big|_{\theta=\theta_{\mathrm{t}}}\right] =0$. 
As the FI is bounded, the variance of $   \partial_{\theta}\log p\left(K,M;\theta,\check{\theta}_{\rm p},\check{\gamma}\right)\Big|_{\theta=\theta_{\mathrm{t}}}$ is:
$E\left[ \left(\partial_{\theta} \log p\left(K,M;\theta,\check{\theta}_{\rm p},\check{\gamma}\right)\right)^2\Big|_{\theta=\theta_{\mathrm{t}}} \right] = \mathcal{I}_{\theta_{\mathrm{t}},\check{\theta}_{\rm p},\check{\gamma}}< \infty$. 
By the central limit theorem,
\begin{align}
\frac{1}{\sqrt{n^{\prime}}} \sum_{i}\partial_{\theta}\log p\left(K,M;\theta,\check{\theta}_{\rm p},\check{\gamma}\right)\Big|_{\theta=\theta_{\mathrm{t}}} \overset{d}{\to} \mathcal{N} \left(0,\mathcal{I}_{\theta_{\mathrm{t}},\check{\theta}_{\rm p},\check{\gamma}}\right). \label{eq:numerator converge}
\end{align}
Consider the denominator in \eqref{eq: ratio}. By the weak law of large numbers (WLLN),
\begin{align}
\frac{1}{n^{\prime}}\sum_{i} \partial^2_{\theta}\log p\left(K_i,M_i;\theta,\check{\theta}_{\rm p},\check{\gamma}\right)\Big|_{\theta=\theta_{\mathrm{t}}} \overset{p}{\to} \mathcal{I}_{\theta_{\mathrm{t}},\check{\theta}_{\rm p},\check{\gamma}}. \label{eq:d weak}
\end{align}
Thus, for $\epsilon_5>0$,
\begin{widetext}
\begin{align}
   &\Pr\left\{ \left|\frac{1}{n^{\prime}}\sum_{i} \partial^2_{\theta}\log p\left(K_i,M_i;\theta,\check{\theta}_{\rm p},\check{\gamma}\right)\Big|_{\theta=\bar{\Theta}} 
 -\mathcal{I}_{\theta_{\mathrm{t}},\check{\theta}_{\rm p},\check{\gamma}}\right| > \epsilon_5 \right\}   \nonumber \\
	&\qquad=\Pr\bigg\{ \bigg|\frac{1}{n^{\prime}}\sum_{i} \partial^2_{\theta}\log p\left(K_i,M_i;\theta,\check{\theta}_{\rm p},\check{\gamma}\right)\Big|_{\theta=\bar{\Theta}} - \partial^2_{\theta}\log p\left(K_i,M_i;\theta,\check{\theta}_{\rm p},\check{\gamma}\right)\Big|_{\theta=\theta_{\mathrm{t}}} \\
    &\qquad\phantom{=\Pr\bigg\{ \bigg|\frac{1}{n^{\prime}}\sum_{i}}+\partial^2_{\theta}\log p\left(K_i,M_i;\theta,\check{\theta}_{\rm p},\check{\gamma}\right)\Big|_{\theta=\theta_{\mathrm{t}}} -\mathcal{I}_{\theta_{\mathrm{t}},\check{\theta}_{\rm p},\check{\gamma}} \bigg| > \epsilon_5 \bigg\}\nonumber\\
	&\qquad\le\Pr\bigg\{ \bigg|\frac{1}{n^{\prime}}\sum_{i} \partial^2_{\theta}\log p\left(K_i,M_i;\theta,\check{\theta}_{\rm p},\check{\gamma}\right)\Big|_{\theta=\bar{\Theta}} -  \frac{1}{n^{\prime}}\sum_{i} \partial^2_{\theta}\log p\left(K_i,M_i;\theta,\check{\theta}_{\rm p},\check{\gamma}\right)\Big|_{\theta=\theta_{\mathrm{t}}}\bigg|> \frac{\epsilon_5}{2}\bigg\}\nonumber\\
&\qquad\phantom{\leq}+\Pr\left\{ \bigg| \frac{1}{n^{\prime}}\sum_{i} \partial^2_{\theta}\log p\left(K_i,M_i;\theta,\check{\theta}_{\rm p},\check{\gamma}\right)\Big|_{\theta=\theta_{\mathrm{t}}}- \mathcal{I}_{\theta_{\mathrm{t}},\check{\theta}_{\rm p},\check{\gamma}}\bigg| > \frac{\epsilon_5}{2} \right\}.  \label{eq: int d converge}
\end{align}
\end{widetext}
The first term in \eqref{eq: int d converge} is zero by the continuity of $ \frac{1}{n^{\prime}}\sum_{i} \partial^2_{\theta}\log p(k_i,m_i;\theta,\check{\theta}_{\rm p},\check{\gamma}) $ over $\theta$, \eqref{eq: bar converge}, and continuous mapping theorem \cite[Th.~25.7]{billingsley95measure}. The second term in \eqref{eq: int d converge} is zero by $ \eqref{eq:d weak}$.
Therefore,  
\begin{align}
\frac{1}{n^{\prime}}\sum_{i} \partial^2_{\theta}\log p(K_i,M_i;\theta,\check{\theta}_{\rm p},\check{\gamma})\Big|_{\theta=\bar{\Theta}} \overset{p}{\to} \mathcal{I}_{\theta_{\mathrm{t}},\check{\theta}_{\rm p},\check{\gamma}}. \label{eq: denominator converge}
\end{align}
Combining \eqref{eq:numerator converge} and \eqref{eq: denominator converge} using Slutsky's theorem yields
\begin{align}
\sqrt{n^{\prime}}\left( \check{\Theta}_{\rm r}\left(\check{\theta}_{\rm p},\check{\gamma}\right) - \theta_{\mathrm{t}} \right) \overset{d}{\to} \mathcal{N} \left(0,\mathcal{I}_{\theta_{\mathrm{t}},\check{\theta}_{\rm p},\check{\gamma}}^{-1}\right). \label{eq: normality}
\end{align}

\section{Regularity Conditions for Asymptotic Normality}
\label{app: normality}

We show \eqref{eq:interchange 1} and \eqref{eq:interchange 2} by using Lemma \ref{lemma: interchange} on $ p(k,m;\theta,\check{\theta}_{p}, \check{\gamma}) $, with $ \mathcal{X} $ being $ \Gamma $, and $ \mathcal{W} $ being the product counting measure space of $ K \times M $. Condition \ref{item:derivative lemma measurability condition} of Lemma \ref{lemma: interchange} holds since $ p(k,m;\theta,\check{\theta}_{p},\check{\gamma}) $ and $ \partial_{\theta} p(k,m;\theta,\check{\theta}_{p},\check{\gamma}) $ are continuous over $\theta$ and, hence, measurable \cite[Cor.~2.2]{folland1999real}.  
We defer treatment of condition \ref{item:derivative lemma absolute continuity condition} to the end of this appendix, instead focusing on condition \ref{item:derivative lemma integral condition} next.
For any $[\theta_1,\theta_2]\subseteq\Gamma\equiv\left[\theta^{\prime\prime},1\right]$, we need:
\begin{widetext}
\begin{align}
\int_{\theta_1}^{\theta_2}    \sum_{k,m}\left|  \partial_{\theta} p(k,m;\theta,\check{\theta}_{\rm p},\check{\gamma})   \right| \dif\theta & \le \int_{\theta^{\prime\prime}}^{1}   \sum_{k,m}\left|  \partial_{\theta} p(k,m;\theta,\check{\theta}_{\rm p},\check{\gamma})   \right| \dif\theta  \leq\max_{\theta\in\Gamma}\sum_{k,m}\left|  \partial_{\theta} p(k,m;\theta,\check{\theta}_{\rm p},\check{\gamma})   \right| <\infty \label{eq:int sum d1} \\
\int_{\theta_1}^{\theta_2}    \sum_{k,m}\left|  \partial^{2}_{\theta} p(k,m;\theta,\check{\theta}_{\rm p},\check{\gamma})   \right| \dif\theta &\le \int_{\theta^{\prime\prime}}^{1}   \sum_{k,m}\left|  \partial^{2}_{\theta} p(k,m;\theta,\check{\theta}_{\rm p},\check{\gamma})   \right| \dif\theta  \leq\max_{\theta\in\Gamma} \sum_{k,m}\left|  \partial^{2}_{\theta} p(k,m;\theta,\check{\theta}_{\rm p},\check{\gamma})   \right|<\infty\label{eq:int sum d2}
\end{align}
\end{widetext}

In order to obtain the bounds in \eqref{eq:int sum d1} and \eqref{eq:int sum d2}, we first note that the output state $ \hat{\sigma}_{IR} =\hat{U}_{R}(\gamma_t) \hat{S}_{IR}^{\dagger}(\zeta) \hat{\sigma}_{  IR}^{d} \hat{S}_{IR}(\zeta)\hat{U}_{R}^{\dagger}(\gamma_t)  $ is a phase-shifted two-mode squeezed thermal state, where $ \hat{\sigma}_{IR}^{d} = \sum_{k,m} r_{k,m} \ket{k,m}\bra{k,m} $ and $ r_{k,m} $ is defined in \eqref{eq:rst}.
Thus:
\begin{align}
\hat{\sigma}_{  IR}^{-1} & = \hat{U}_{R}(\gamma_t)\hat{S}_{IR}^{\dagger}(\zeta) \left(\hat{\sigma}_{ IR}^{d} \right)^{-1}  \hat{S}_{IR}(\zeta)\hat{U}_{R}^{\dagger}(\gamma_t)\\
\hat{\sigma}_{  IR}^{\frac{1}{2}} & =  \hat{U}_{R}(\gamma_t)\hat{S}_{IR}^{\dagger}(\zeta) \left(\hat{\sigma}_{  IR}^{d} \right)^{\frac{1}{2}}  \hat{S}_{IR}(\zeta)\hat{U}_{R}^{\dagger}(\gamma_t).
\end{align}
We use the following traces of the combinations of $\hat{\sigma}_{IR}$ with creation and annihilation operators in subsequent calculations:
\begin{align}
\trace\left\{ \hat{\sigma}_{IR}\hat{a}_{R}^{\dagger}\hat{a}_{R}\right\} &=   \mu^2 N_{1} + \nu^2(1+N_{2})\label{eq:term ad a}\\
\trace\left\{ \hat{\sigma}_{IR}\hat{a}_{R}\hat{a}_{R}^{\dagger}\right\} &= \mu^2 N_{1} + \nu^2(1+N_{2})+1\label{eq:term a ad}\\
\trace\left\{ \hat{\sigma}_{IR}\hat{a}_{R}^{\dagger}\hat{a}_{R}^{\dagger}\right\} &=   
\trace\left\{ \hat{\sigma}_{IR}\hat{a}_{R}\hat{a}_{R}\right\} = 0\label{eq:expectation a a}\\
\trace\left\{ \hat{\sigma}_{IR} \hat{a}_{R} \hat{a}_{R}^{\dagger}\hat{a}_{R}\hat{a}_{R}^{\dagger} \right\}&=  1+3N_{1}\mu^2+3(1+N_{2})\nu^2\nonumber\\
&\quad+2(N_{1}\mu^2 +(1+N_{2})\nu^2)^{2}\label{eq:term aadaad}\\
\trace\left\{ \hat{\sigma}_{IR} \hat{a}_{R}^{\dagger}\hat{a}_{R}\hat{a}_{R}^{\dagger}\hat{a}_{R} \right\}&= N_{1}\mu^2+(1+N_{2})\nu^2\nonumber\\
&\quad+2(N_{1}\mu^2+(1+N_{2})\nu^2)^{2},\label{eq:term adaada}
\end{align}
where $\mu = \cosh \zeta$ and $\nu = -\sinh\zeta$, with $\zeta$, $N_1$, and $N_2$ being the functions of $\theta_{\rm t}$, $\bar{n}_{\rm S}$, and $\bar{n}_{\rm B}$ given in \cite[Eqs.~(56), (59), and (60)]{gong22losssensing}.
We obtain \eqref{eq:term ad a}--\eqref{eq:term adaada} from $\hat{\sigma}_{IR}$ being a Gaussian state \cite{weedbrook12gaussianQIrmp}. 
The returned probe state evolving in thermal bath is characterized by the Lindblad master equation \cite[Ch.~4]{ferraro05gaussian}:
\begin{align}
\partial_{\theta} \hat{\sigma}_{IR}= -\frac{1}{2\theta}\left [ (\bar{n}_{\rm B}+1) \hat{\mathcal{L}}_{R}[\hat{a}]+\bar{n}_{\rm B}\hat{\mathcal{L}}_{R}[\hat{a}^{\dagger}]\right ] \hat{\sigma}_{IR}, \label{eq:lindblad master equation}
\end{align}
where 
\begin{align}
\hat{\mathcal{L}}_{R}[\hat{a}] \hat{\sigma}_{IR}&= 2 \hat{a}_{R}\hat{\sigma}_{IR} \hat{a}_{R}^{\dagger} - \hat{a}_{R}^{\dagger}\hat{a}_{R}\hat{\sigma}_{IR} - \hat{\sigma}_{IR}\hat{a}_{R}^{\dagger}\hat{a}_{R}\\
\hat{\mathcal{L}}_{R}[\hat{a}^{\dagger}] \hat{\sigma}_{IR}&= 2 \hat{a}_{R}^{\dagger}\hat{\sigma}_{IR} \hat{a}_{R} \!- \hat{a}_{R}\hat{a}_{R}^{\dagger}\hat{\sigma}_{IR} - \hat{\sigma}_{IR}\hat{a}_{R}\hat{a}_{R}^{\dagger}.
\end{align}

Now, we  obtain \eqref{eq:int sum d1}.
Using \eqref{eq:lindblad master equation} and triangle inequality,
\begin{widetext}
\begin{align}
   \sum_{k,m} \left|\partial_{\theta}p(k,m;\theta,\check{\theta}_{\rm p},\check{\gamma})\right| & =\sum_{k,m} \left|\bra{k,m} \hat{S}_{IR}^{\dagger}(\omega(\check{\theta}_{\rm p}))\hat{U}_{R}(\gamma_t-\check{\gamma})\partial_{\theta}\hat{\sigma}_{IR}(\theta)\hat{U}_{R}^{\dagger}(\gamma_t-\check{\gamma})\hat{S}_{IR}(\omega(\check{\theta}_{\rm p})) \ket{k,m}\right|\nonumber\\
	&\le \frac{\bar{n}_{\rm B}+1}{2\theta}\sum_{k,m} \left|  \bra{k,m} \hat{S}_{IR}^{\dagger}(\omega(\check{\theta}_{\rm p})) \hat{U}_{R}(\gamma_t-\check{\gamma}) 2 \hat{a}_{R}\hat{\sigma}_{IR} \hat{a}_{R}^{\dagger}\hat{U}_{R}^{\dagger}(\gamma_t-\check{\gamma})  \hat{S}_{IR}(\omega(\check{\theta}_{\rm p})) \ket{k,m} \right| \nonumber\\
	&\phantom{=} +  \frac{\bar{n}_{\rm B}+1}{2\theta}\sum_{k,m} \left| \bra{k,m} \hat{S}_{IR}^{\dagger}(\omega(\check{\theta}_{\rm p})) \hat{U}_{R}(\gamma_t-\check{\gamma})\left( \hat{a}_{R}^{\dagger}\hat{a}_{R}\hat{\sigma}_{IR}   +    \hat{\sigma}_{IR}\hat{a}_{R}^{\dagger}\hat{a}_{R}\right) \right.\nonumber \\
 &\left.\vphantom{  \bra{k,m}  \hat{S}_{IR} } \qquad \qquad \qquad \qquad \times\hat{U}_{R}^{\dagger}(\gamma_t-\check{\gamma})  \hat{S}_{IR}(\omega(\check{\theta}_{\rm p})) \ket{k,m} \right |\nonumber\\
	&\phantom{=} + \frac{\bar{n}_{\rm B}}{2\theta} \sum_{k,m}\left| \bra{k,m}  \hat{S}_{IR}^{\dagger}(\omega(\check{\theta}_{\rm p})) \hat{U}_{R}(\gamma_t-\check{\gamma})  \left( \hat{a}_{R}\hat{a}_{R}^{\dagger}\hat{\sigma}_{IR}  +      \hat{\sigma}_{IR}\hat{a}_{R}\hat{a}_{R}^{\dagger}\right)\hat{U}_{R}^{\dagger}(\gamma_t-\check{\gamma})  \hat{S}_{IR}(\omega(\check{\theta}_{\rm p})) \ket{k,m} \right| \nonumber \\
&\phantom{=}+\frac{\bar{n}_{\rm B}}{2\theta}\sum_{k,m} \left| \bra{k,m} \hat{S}_{IR}^{\dagger}(\omega(\check{\theta}_{\rm p})) \hat{U}_{R}(\gamma_t-\check{\gamma}) 2 \hat{a}_{R}^{\dagger}\hat{\sigma}_{IR} \hat{a}_{R}  \hat{U}_{R}^{\dagger}(\gamma_t-\check{\gamma}) \hat{S}_{IR}(\omega(\check{\theta}_{\rm p}))\ket{k,m}\right|. \label{eq:f up}
\end{align}  
\end{widetext}

Since $ \hat{\sigma}_{IR} $ is positive-definite, the first and third terms in \eqref{eq:f up} are bounded per \eqref{eq:term ad a} and \eqref{eq:term a ad}:
\begin{widetext}
\begin{align}
  &\frac{\bar{n}_{\rm B}+1}{2\theta}\sum_{k,m} \left|  \bra{k,m} \hat{S}_{IR}^{\dagger}(\omega(\check{\theta}_{\rm p})) \hat{U}_{R}(\gamma_t-\check{\gamma})  2 \hat{a}_{R}\hat{\sigma}_{IR} \hat{a}_{R}^{\dagger}\hat{U}_{R}^{\dagger}(\gamma_t-\check{\gamma})  \hat{S}_{IR}(\omega(\check{\theta}_{\rm p})) \ket{k,m} \right| \nonumber \\
	&=  \frac{\bar{n}_{\rm B}+1}{2\theta}\sum_{k,m}    \bra{k,m} \hat{S}_{IR}^{\dagger} (\omega(\check{\theta}_{\rm p})) \hat{U}_{R}(\gamma_t-\check{\gamma})  2 \hat{a}_{R}\hat{\sigma}_{IR} \hat{a}_{R}^\dagger\hat{U}_{R}^{\dagger}(\gamma_t-\check{\gamma})  \hat{S}_{IR}(\omega(\check{\theta}_{\rm p})) \ket{k,m} = \frac{\bar{n}_{\rm B}+1}{\theta}\trace\left\{ \hat{a}_{R}^{\dagger} \hat{a}_{R}\hat{\sigma}_{IR}  \right \}<\infty\\
  &\frac{\bar{n}_{\rm B}}{2\theta}\sum_{k,m} \left| \bra{k,m} \hat{S}_{IR}^{\dagger} (\omega(\check{\theta}_{\rm p})) \hat{U}_{R}(\gamma_t-\check{\gamma})  2 \hat{a}_{R}^{\dagger}\hat{\sigma}_{IR} \hat{a}_{R}\hat{U}_{R}^{\dagger}(\gamma_t-\check{\gamma})  \hat{S}_{IR} (\omega(\check{\theta}_{\rm p}))\ket{k,m}\right| \nonumber\\
	&=  \frac{\bar{n}_{\rm B}}{2\theta}\sum_{k,m}   \bra{k,m} \hat{S}_{IR}^{\dagger} (\omega(\check{\theta}_{\rm p})) \hat{U}_{R}(\gamma_t-\check{\gamma})  2 \hat{a}_{R}^{\dagger}\hat{\sigma}_{IR} \hat{a}_{R} \hat{U}_{R}^{\dagger}(\gamma_t-\check{\gamma})  \hat{S}_{IR} (\omega(\check{\theta}_{\rm p}))\ket{k,m}  = \frac{\bar{n}_{\rm B}}{\theta}\trace\left\{ \hat{a}_{R} \hat{a}_{R}^{\dagger}\hat{\sigma}_{IR}  \right \}<\infty.
\end{align}
\end{widetext}

The second term in \eqref{eq:f up} is the sum of the diagonal terms of an operator. Consider the following upper bound:
\begin{align}
\sum_{i} |a_{ii}| = \trace\left\{ \hat{A}\hat{B} \right\}    \le \sum_{i} |\lambda_i| = \trace\left\{  \sqrt{   \hat{A} \hat{A}^{\dagger} }\right\} \label{eq:von neu inequality},
\end{align}
where $ \hat{A} $ is a Hermitian operator with $ ij $-th entry $ a_{ij} $, $ \hat{B} $ is a diagonal operator with diagonal element $ b_i $ such that $ a_{ii}b_i = |a_{ii}| $, $ \lambda_i $ is the eigenvalue of $ \hat{A} $, and the inequality in \eqref{eq:von neu inequality} is by von Neumann trace inequality \cite{mirsky1975trace}. Thus, the second term in \eqref{eq:f up} is bounded as follows:
\begin{widetext}
\begin{align}
    &\sum_{k,m}\Big| \bra{k,m}  \hat{S}_{IR}^{\dagger}(\omega(\check{\theta}_{\rm p}))  \hat{U}_{R}(\gamma_t-\check{\gamma})  \left( \hat{a}_{R}^{\dagger}\hat{a}_{R}\hat{\sigma}_{IR}  +      \hat{\sigma}_{IR}\hat{a}_{R}^{\dagger}\hat{a}_{R}\right) \hat{U}_{R}^{\dagger}(\gamma_t-\check{\gamma}) \hat{S}_{IR}(\omega(\check{\theta}_{\rm p})) \ket{k,m} \Big| \nonumber\\
	&\qquad\le  \trace\left\{  \left(  \left( \hat{a}_{R}^{\dagger}\hat{a}_{R}\hat{\sigma}_{IR} +    \hat{\sigma}_{IR}\hat{a}_{R}^{\dagger}\hat{a}_{R}\right)\left( \hat{a}_{R}^{\dagger}\hat{a}_{R}\hat{\sigma}_{IR} +    \hat{\sigma}_{IR}\hat{a}_{R}^{\dagger}\hat{a}_{R}\right)  \right)^{\frac{1}{2}} \right\} \label{eq:fd v u}\\
	&\qquad\le \left( \trace\left\{   \left( \hat{a}_{R}^{\dagger}\hat{a}_{R}\hat{\sigma}_{IR} +    \hat{\sigma}_{IR}\hat{a}_{R}^{\dagger}\hat{a}_{R}\right) \left( \hat{a}_{R}^{\dagger}\hat{a}_{R}\hat{\sigma}_{IR} +    \hat{\sigma}_{IR}\hat{a}_{R}^{\dagger}\hat{a}_{R}\right) \hat{\sigma}^{-1}_{IR}  \right\}\right)^{\frac{1}{2}}  \label{eq:fd u}\\
	& \qquad=  \left( 3 \trace\left\{ \hat{\sigma}_{IR}\hat{a}_{R}^{\dagger}\hat{a}_{R} \hat{a}_{R}^{\dagger}\hat{a}_{R}  \right\} +\trace\left\{\hat{a}_{R}^{\dagger}\hat{a}_{R} \hat{\sigma}_{IR} \hat{\sigma}_{IR}\hat{a}_{R}^{\dagger}\hat{a}_{R}    \hat{\sigma}_{IR}^{-1}   \right\}\right)^{\frac{1}{2}} <\infty \label{eq:second u},
\end{align}
where  \eqref{eq:fd v u} is by \eqref{eq:von neu inequality}, and \eqref{eq:fd u} is by Holder's inequality and $ \sqrt{\trace\{\hat{\sigma}_{IR}\} } = 1  $.  
The first term in \eqref{eq:second u} is finite by \eqref{eq:term adaada}. The second term in \eqref{eq:second u} is also finite:
\begin{align}
      &\trace\left\{  \hat{a}_{R}^{\dagger}\hat{a}_{R} \hat{\sigma}_{IR} \hat{\sigma}_{IR}\hat{a}_{R}^{\dagger}\hat{a}_{R}   \hat{\sigma}_{IR}^{-1}   \right\} =\trace\left\{ \hat{b}_{R}^{\dagger}\hat{b}_{R} \left(\hat{\sigma}_{IR}^{d} \right)^{2}  \hat{b}_{R}^{\dagger}\hat{b}_{R}   \left(\hat{\sigma}_{IR}^{d} \right)^{-1}    \right \}= \sum_{k,m} \sum_{k^{\prime}m^{\prime}}r^{2}_{k,m}r^{-1}_{k^{\prime}m^{\prime}} \left|\bra{k^{\prime}m^{\prime}} \hat{c}_{IR} \ket{k,m} \right|^2,\label{eq:sumation}
\end{align}
where $ \hat{b}_{R} = \hat{S}_{IR}^{\dagger}(\zeta)  \hat{a}_{R}  \hat{S}_{IR}(\zeta) = \mu\hat{a}_{R} +\nu\hat{a}_{I}^{\dagger} $, and $\hat{c}_{IR}\equiv \mu^2\hat{a}_{R}^{\dagger}\hat{a}_{R}+\mu\nu\left( \hat{a}_{R}^{\dagger}\hat{a}_{I}^{\dagger} + \hat{a}_{R} \hat{a}_{I}  \right) +  \nu^2 \hat{a}_{I}\hat{a}_{I}^{\dagger}$.

When $k^{\prime}=k$ and $m^{\prime}=m$ in the inner sum of \eqref{eq:sumation}, we have:
\begin{align}
   &\sum_{k,m}r_{k,m}   \left( \bra{k m } \mu^2\hat{a}_{R}^{\dagger}\hat{a}_{R}  +  \nu^2 \hat{a}_{I}\hat{a}_{I}^{\dagger} \ket{k,m} \right)^2  \nonumber \\
	&\qquad=  \sum_{k,m}r_{k,m}  \left(\mu^2k  +  \nu^2 (m+1) \right)^2= N_{1}(1+2N_{1})\mu^4+2N_{1}(1+2N_{1})\mu^2\nu^2 + (1+N_{2})(1+2N_{2})\nu^4.\label{eq:second upper 1}
\end{align}
When $k^{\prime}=k+1$ and $m^{\prime}=m+1$, we have:
\begin{align}
&\sum_{k,m}r_{k,m}^{2} r_{k+1,m+1}^{-1}  \bra{k+1, m+1 } \mu\nu \hat{a}_{R}^{\dagger}\hat{a}_{I}^{\dagger}   \ket{ k, m } \bra{ k, m } \mu\nu \hat{a}_{R} \hat{a}_{I}   \ket{k+1, m+1 } \nonumber \\
&\qquad=\sum_{k,m}r_{k,m}^{2} r_{k+1,m+1}^{-1}  \mu^2\nu^2   (k+1)(m+1)=  \frac{(1+N_{1})^{2}(1+N_{2})^{2}\mu^2\nu^2}{N_1N_2}. \label{eq:second upper 2}
\end{align}
When $k^{\prime}=k-1$ and $m^{\prime}=m-1$, we have:
\begin{align}
  &\sum_{k^{\prime}m^{\prime}}r_{k^{\prime}+1,m^{\prime}+1}^{2} r_{k^{\prime},m^{\prime}}^{-1}  \bra{k^{\prime}, m^{\prime} } (\mu\nu \hat{a}_{R} \hat{a}_{I}    ) \ket{ k^{\prime}+1, m^{\prime}+1 }\bra{ k^{\prime}+1, m^{\prime}+1 }(\mu\nu \hat{a}_{R}^{\dagger} \hat{a}_{I}^{\dagger} )\ket{ k^{\prime}, m^{\prime} }\nonumber\\
	&\qquad=  \sum_{k,m}r_{k^{\prime}+1,m^{\prime}+1}^{2} r_{k^{\prime},m^{\prime}}^{-1}  \mu^2\nu^2   (k+1)(m+1)= \frac{N_1^{2}N_2^{2}\mu^2\nu^2}{(1+N_1)(1+N_2)}. \label{eq:second upper 3}
\end{align}
\end{widetext}
Since other combinations of values for $k^\prime$ and $m^\prime$ result in zeros, \eqref{eq:sumation} is bounded. 
Hence, the second term in \eqref{eq:f up} is bounded. The third term in \eqref{eq:f up} is bounded similarly: we adapt the use of Holder's inequality in steps leading to \eqref{eq:fd u}. This yields the following bound:
\begin{widetext}
\begin{align}
& \trace\left\{   \left( \hat{a}_{R}\hat{a}_{R}^{\dagger}\hat{\sigma}_{IR} +    \hat{\sigma}_{IR}\hat{a}_{R}\hat{a}_{R}^{\dagger}\right)   \left( \hat{a}_{R}\hat{a}_{R}^{\dagger}\hat{\sigma}_{IR} +    \hat{\sigma}_{IR}\hat{a}_{R}\hat{a}_{R}^{\dagger}\right) \hat{\sigma}^{-1}_{IR}  \right\}\nonumber \\
  &\qquad =\trace\left\{\hat{a}_{R}^{\dagger}\hat{a}_{R} \hat{\sigma}_{IR} \hat{\sigma}_{IR}\hat{a}_{R}^{\dagger}\hat{a}_{R} \hat{\sigma}_{IR}^{-1}   \right\}+  3 \trace\left\{ \hat{\sigma}_{IR}\hat{a}_{R}\hat{a}_{R}^{\dagger} \hat{a}_{R}\hat{a}_{R}^{\dagger}  \right\}   +2\trace\left\{  \hat{a}_{R}^{\dagger}\hat{a}_{R}  \hat{\sigma}_{IR}        \right\} +1<\infty,  \label{eq:fourth upper}
\end{align}
\end{widetext}
where \eqref{eq:fourth upper} is by  \eqref{eq:term ad a}, \eqref{eq:term aadaad}, and \eqref{eq:sumation}-\eqref{eq:second upper 3}. Since all terms in \eqref{eq:f up} are finite, there exists $C_2>0 $ such that, for all $ \theta\in\Gamma $,
\begin{align}
\sum_{k,m}\left|  \partial_{\theta} p(k,m;\theta,\check{\theta}_{\rm p})   \right| < C_2.\label{eq:upper d1}
\end{align}
An upper bound on $ \sum_{k,m}\left|  \partial_{\theta}^2 p(k,m;\theta,\check{\theta}_{\rm p})   \right| $ is obtained similarly to \eqref{eq:upper d1}; we omit the lengthy derivation. Hence, there exists $ C_3>0 $ such that, for all $ \theta\in\Gamma $,
\begin{align}
\sum_{k,m}\left|  \partial_{\theta}^2 p(k,m;\theta,\check{\theta}_{\rm p})   \right| < C_3 \label{eq:upper d2}.
\end{align}
Finally, to show that condition \ref{item:derivative lemma absolute continuity condition} of Lemma \ref{lemma: interchange} holds, we first point out integrability: $\sum_{k,m} |p(k,m;\theta,\check{\theta}_{p},\check{\gamma})| =1$ since $p(k,m;\theta,\check{\theta}_{p},\check{\gamma})$ is a p.m.f.~and $ \sum_{k,m} |\partial_{\theta}p(k,m;\theta,\check{\theta}_{p},\check{\gamma})| <\infty $ by \eqref{eq:upper d1}.
Then, for all compact intervals $ [\theta_1, \theta_2] \subseteq \Gamma $, $ p(k,m;\theta,\check{\theta}_{p},\check{\gamma}) $ and $ \partial_{\theta} p(k,m;\theta,\check{\theta}_{p},\check{\gamma}) $ are absolutely continuous by FTCL since 
\begin{widetext}
\begin{align} 
p(k,m;\theta_2,\check{\theta}_{p},\check{\gamma}) - p(k,m;\theta_1,\check{\theta}_{p},\check{\gamma})  &= \int_{\theta_1}^{\theta_2}\partial_{\theta}p(k,m;\theta,\check{\theta}_{p},\check{\gamma}) \dif\theta \\
\partial_{\theta}p(k,m;\theta_2,\check{\theta}_{p},\check{\gamma}) - \partial_{\theta}p(k,m;\theta_1,\check{\theta}_{p},\check{\gamma}) &= \int_{\theta_1}^{\theta_2} \partial^{2}_{\theta}p(k,m;\theta,\check{\theta}_{p},\check{\gamma}) \dif\theta,
\end{align}

and $ \partial_{\theta}p(k,m;\theta,\check{\theta}_{p},\check{\gamma}) $ and $ \partial^2_{\theta}p(k,m;\theta,\check{\theta}_{p},\check{\gamma})$ are integrable by \eqref{eq:upper d1} and \eqref{eq:upper d2}. 
\end{widetext}

\end{document}